\documentclass{aa}
\usepackage{graphicx}
\usepackage{natbib}
\def\pdb{Plateau de Bure}

\def\oo{ON--OFF}
\def\uvc{$uv$-coverage}

\begin{document}
   \title{Interferometric mapping of the 3.3-mm continuum emission of comet 17P/Holmes after its 2007 outburst\thanks{Based on observations carried out with
the IRAM Plateau de Bure Interferometer. IRAM is supported by INSU/CNRS (France), MPG (Germany) and IGN (Spain).}}

   \author{J. Boissier
          \inst{1,2}
          \and
          D. Bockel\'ee-Morvan\inst{3}
      \and
          N. Biver\inst{3}
          \and
          J. Crovisier\inst{3}
      \and
              E. Lellouch\inst{3}
      \and
          R. Moreno\inst{3}
       \and
          V. Zakharov\inst{3}   
          }
   \institute{Istituto di Radioastronomia, INAF, Via Gobetti 101, Bologna, Italy\\
              \email{boissier@ira.inaf.it}
              \and
         ESO, Karl Schwarzschild St. 2, 85748 Garching, Muenchen, Germany    
         \and
             LESIA, Observatoire de Paris, CNRS, UPMC, Universit\'e Paris-Diderot, 
               5 place Jules Janssen, 92195 Meudon, France\\}
   \date{revised version}
  \abstract
  {Comet 17P/Holmes underwent a dramatic outburst in October 2007,  caused by the 
   sudden fragmentation of its nucleus and the production of a large quantity of grains scattering sunlight.}
   {We report on 90 GHz continuum observations carried out with the IRAM \pdb{} interferometer on 27.1 and 28.2 October 2007 UT, i.e., 4--5 days after the outburst. These observations
probed the thermal radiation of large dust particles, and therefore provide the best constraints on the mass in the ejecta debris.  }
   {The thermal emission of the debris was modelled and coupled to a time-dependent description of their expansion after the outburst. The analysis was performed in the Fourier plane. Visibilities were computed for the two observing dates and compared to the data to measure their velocity and mass.  Optical data and 250-GHz continuum measurements published in the literature were used to further constrain the dust kinematics and size distribution.    }
   {Two distinct dust components in terms of kinematic properties are identified in the data. The large-velocity component, with typical velocities $V_0$ of 50--100 m s$^{-1}$ for 1 mm particles,  displays a steep size distribution with a size index estimated to  $q$ = --3.7 ($\pm$0.1), assuming a minimum grain size of 0.1 $\mu$m. It corresponds to the fast expanding shell observed in optical images. The slowly-moving "core" component ($V_0$ = 7--9 m s$^{-1}$)  detected near the nucleus has a size index $|q|$ $<$ 3.4 and contains a higher proportion of large particles than the shell. The dust mass in the core is in the range 0.1--1 that of the shell.  Using optical constants pertaining to porous grains (50\% porosity) made of astronomical silicates mixed with water ice (48\% in mass), the total dust mass $M_{\rm dust}$ injected by the outburst is estimated to 4--14 $\times$ 10$^{11}$ kg, corresponding to 3--9\% the nucleus mass. }
   {}
 
   \titlerunning{The 3.3-mm continuum emission of comet 17P/Holmes after its 2007 outburst}
   \maketitle
{}

\section{Introduction}

Comet 17P/Holmes is a periodic comet of the Jupiter family that
orbits the Sun with a period $P$ = 6.9 years. It passed perihelion
on 4 May 2007  at 2.05 AU from the Sun. On 24 October 2007, at 2.44 AU
from the Sun and 1.63 AU from the Earth, the comet suddenly
increased in brightness from a total visual magnitude $m_v$ $\sim$
17 to 2.5 to become a naked-eye object for months \citep{Green2007,Sekanina2009}. Comet
17P/Holmes underwent a similar outburst shortly before 6 November
1892, at the time of its discovery, followed by a similar event on
16 January 1893 \citep[see the review of][]{Sekanina2009}. After these events, the comet appearance at large
scales was a bubble-like shape quickly expanding into
interplanetary space \citep[e.g.,][]{montalto2008,Hsieh2010}. The onset of the 2007 outburst occurred
probably near 23.3 October UT \citep{Hsieh2010} with the peak of
optical brightening observed around 25.0 October  UT
\citep[e.g.,][]{Li2011}. These outbursts were likely caused by a
sudden fragmentation of the nucleus, followed by the
production of a large quantity of grains scattering sunlight.
Determining the amount of material that split off from the nucleus
and the size distribution of the particle debris in the 
cloud of dust ejecta is important to constrain the origin of the
fragmentation process. Though sizeable individual fragments
radiating outwards were possibly observed \citep{gail07,steven10}, the huge
cross-sectional area of dust scattering sunlight suggests that the
dust in the comet Holmes 2007 outburst was dominated by small
particles.

The potential of millimetre and submillimetre-wavelength continuum
observations for the study of cometary dust has been demonstrated
\citep[e.g.,][]{jew90,jew92,jewmat99,alten99,dbm2010b}. Such observations
probe the thermal radiation of millimetre-sized dust particles,
and therefore usefully complement optical and infrared
observations, that are sensitive to micrometric particles. By
measuring the radiation from large particles, they are of high
value to measure dust masses. Constraints on the dust properties,
e.g., the size distribution, can be obtained if measurements of
the spectral index of the dust emission are available
\citep{jew90,jew92}.

Single-dish continuum measurements at 1.1 mm wavelength of comet 17P/Holmes 
obtained with the 30-m telescope of the Institut de Radioastronomie millim\'etrique (IRAM) have been presented by \citet{alten2009}.
A short report of observations at 1.3 and 0.8 mm conducted with the
Submillimeter Array was given by \citet{qi2010}.
In this paper, we present interferometric continuum observations
 performed at 3.3 mm wavelength on 27
and 28 October 2007 UT. These observations, carried out with the IRAM Plateau
de Bure interferometer, provided images of the dust coma at $\sim$
6$\arcsec$ angular resolution which corresponds to 7100 km diameter at
the comet distance. The observations and data products are presented in Sects.~\ref{sec:1} and \ref{sec:2}. They were analysed
with a model of dust thermal emission coupled with a time-dependent model of the expansion of the cloud. The modelling approach is described in Sect.~\ref{sec:4} and the data analysis is done in Sect.\ref{sec:5}. A discussion of the results obtained on the properties of the dust ejecta follows in Sect.\ref{sec:6}. A preliminary report of the observations was given by \citet{boissier2008,boissier2009}.

\section{Observations}
\label{sec:1}

\begin{table*}
\small
\begin{center}
\caption{Observation log and flux measurements. }
\label{log}
\begin{tabular}{l c c c c c c c}
\hline
\noalign{\smallskip}
\hline
\noalign{\smallskip}
Date UT & Freq. & $N_A$$^a$ & Beam & $S/T_A$$^b$ & Peak intensity & S/N & $\Delta$O$^c$  \\
(October 2007)& (GHz) & & ($\arcsec \times \arcsec$) & (Jy/K)  & (mJy/beam)  &     & ($\arcsec$) \\
\hline
\noalign{\smallskip}
\multicolumn{7}{l}{\emph{Total datasets} $^d$}\\
26.92--27.29 & 88.6 & 5  &6.88$\times$5.97&23.14 &  $2.38 \pm  0.07$ & 34 & 0.2 \\
28.08--28.33  & 90.6 & 6 &6.44$\times$4.77 &22.4 &  $1.9 \pm 0.1$ & 19  & 0.3  \\
\hline
\noalign{\smallskip}
\multicolumn{7}{l}{\emph{Subsets} $^e$}\\
27.08--27.29 & 88.6 & 5  &7.49$\times$5.89& 23.14 &   $2.45 \pm  0.12$ & 20 & 0.3  \\
28.08--28.29 & 90.6 & 5 &7.19$\times$5.66 &22.4 &  $2.06 \pm 0.14$ & 15   & 0.4  \\
\hline
\noalign{\smallskip}
\end{tabular}
\end{center}

\begin{list}{}{}
\item[$^a$] Number of observing  antennas.
\item[$^b$] Conversion factor between flux and antenna temperature.
\item[$^c$] $\Delta$O is the astrometric precision of the position measured on the map. It is given by the interferometric beam divided by the signal to noise ratio.
\item[$^d$] In this part we take into account all the observations for each day.
\item[$^e$] Here the maps were done with the observations performed in the common UT range of 27 and 28 October.
\end{list}
\end{table*}

\begin{figure}
\resizebox{\hsize}{!}{\includegraphics[angle=0]{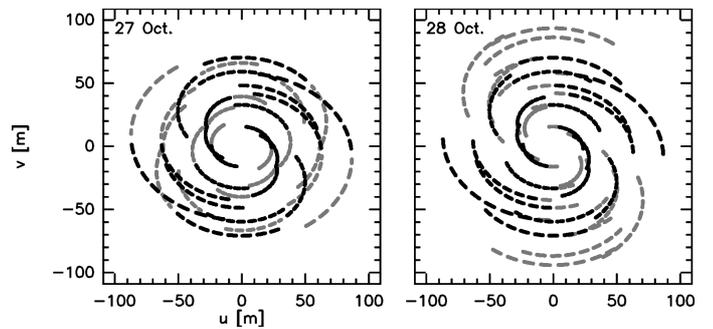}}
    \caption{ \uvc{} for 27 and 28 October. The $(u,v)$ points in black represent the part common to the two dates.
}
\label{fig-uv}
\end{figure}

Observations were undertaken on  26.92--27.29 and
28.08--28.33 October 2007 UT at the IRAM Plateau de Bure interferometer (PdBI) situated in
the French Alps \citep{gui+92}. Therefore, they were performed
soon after the announcement of the comet outburst on  24 October
\citep{Green2007} as a Target of Opportunity program. The comet was at
a geocentric distance $\Delta \sim 1.63 $ AU and a heliocentric
distance $r_{h} \sim 2.45$ AU. The Sun was at a position angle
p.a. = 38--39$^{\circ}$ on the plane of the sky, and the phase
(Sun-Comet-Earth) angle was $\sim$ 16$^{\circ}$. The comet was
tracked using the ephemeris provided by the HORIZONS system
(solution JPL K077/6). At the time of the observations, the six
15-m antennas of the interferometer were set in the compact D
configuration. On 27 October, only five antennas were available.
The baseline lengths projected onto the plane of the sky ranged
from 15 to 100 m. The \uvc{} acquired over the course of observations is presented in Fig.~\ref{fig-uv}.

The observations were performed using the 3-mm dual polarization
(4-GHz bandwidth) receivers tuned to the frequencies of the
$J$(1--0) HCN (88.6 GHz) and $J$(1--0) HNC (90.6 GHz) lines on
27 and 28 October, respectively. The two polarizations were
observed using the 8 units of the \pdb{} correlator: for each
polarization, one unit of 20 MHz bandwidth was centred on the
molecular line, and three 320 MHz units were placed nearby for
measuring the continuum emission. The effective total bandwidth
for the continuum observations was thus about 2 $\times$ 0.9 GHz. The
observing cycle was typically: pointing, focussing,
cross-correlation on a calibration source, 2 min of
autocorrelation (\oo{} measurements, 1 min on source, 
for line observations with the angular resolution of the primary beam) and 30 scans (45 s each) of cross-correlation on
the comet. A log of the observations is presented in
Table~\ref{log}. The  angular resolution of the synthesized beam is $\sim$ 6.4$\arcsec$
and 5.5$\arcsec$ on 27 and 28 October, respectively and the Half Power
Beam Width ($HPBW$) of the primary beam is about 54$\arcsec$ at the
observed frequencies. The analysis of the HCN and HNC data
(cross-correlation, i.e. interferometric mode, and \oo{}
measurements) will be presented in a forthcoming paper.

The data reduction was done using the GILDAS software package \citep{pet05}.
For both datasets (27 and 28 October) the bandpass calibration was carried out observing 3C454.3.
 The instrumental phase and amplitude variations were deduced from observations of the   calibrator   0355+508.
The weather conditions at the \pdb{} were outstanding, and the
resulting phase rms is at most 20$^{\circ}$, with values lower
than 15$^{\circ}$ for most of the antenna pairs. Finally the
absolute flux density scale was determined from measurements of
3C84 and MWC349 fluxes. The consistency (within 2.5\%) of the
calibrator fluxes measured on  27 and 28 October implies an
excellent relative calibration between the two dates. The
uncertainty in the absolute flux calibration is about 15\% for
both days.

\begin{table*}
\begin{center}

\caption{Position of the brightness centre.} \label{pos}
\begin{tabular}{l c c c c c c c}
\hline \noalign{\smallskip} \hline \noalign{\smallskip} Date &
Freq. & UT time$^a$ & \multicolumn{2}{c}{Ephemeris Position
(C)$^{b, c}$} & \multicolumn{2}{c
}{Observed Position (O)$^b$} &(O-C)$^d$ \\
&&&\multicolumn{2}{c}{--------------------------------} & \multicolumn{2}{c}{--------------------------------} & \\
        &  & &RA & Dec & RA & Dec &(RA, Dec) \\
(2007) & (GHz) & (h) & (h:min:s) & ($^{\circ}$:$\arcmin$:$\arcsec$) &  (h:min:s) & ($^{\circ}$:$\arcmin$:$\arcsec$) &($\arcsec$,$\arcsec$)  \\
\hline
\noalign{\smallskip}
\multicolumn{8}{l}{\emph{Total datasets}}\\
27 October & 88.6 & 7.0000 & 03:51:30.418 & 50:19:16.48 & 03:51:30.418 & 50:19:16.76 &(+0.00,+0.28) \\
28 October & 90.6 & 8.0000 & 03:50:30.701 & 50:23:04.37 & 03:50:30.733 & 50:23:04.66 &(+0.31,+0.29) \\
\hline
\noalign{\smallskip}
\multicolumn{8}{l}{\emph{Subsets}}\\
27 October & 88.6 & 7.0000 & 03:51:30.418 & 50:19:16.48 & 03:51:30.436 & 50:19:16.59 &(+0.17,+0.11) \\
28 October & 90.6 & 6.0000 &   03:50:35.552 & 50:22:46.94  & 03:50:35.569 & 50:22:47.26 &(+0.16,+0.31) \\
\hline
\noalign{\smallskip}
\end{tabular}
\end{center}
\begin{list}{}{}
\item[$^a$] Reference time for the position in RA, Dec of the
comet. \item[$^b$] The coordinate system is apparent positions,
geocentric frame. \item[$^c$] Ephemeris computed from the HORIZONS
system, solution JPL K077/15 based on 3265 astrometric data from
1964 July 16 to 2008 April 22.
 \item[$^b$] Offset between the peak of the continuum emission and
the position given by the ephemeris.
\end{list}
\end{table*}

\begin{figure}
\resizebox{\hsize}{!}{\includegraphics[angle=0]{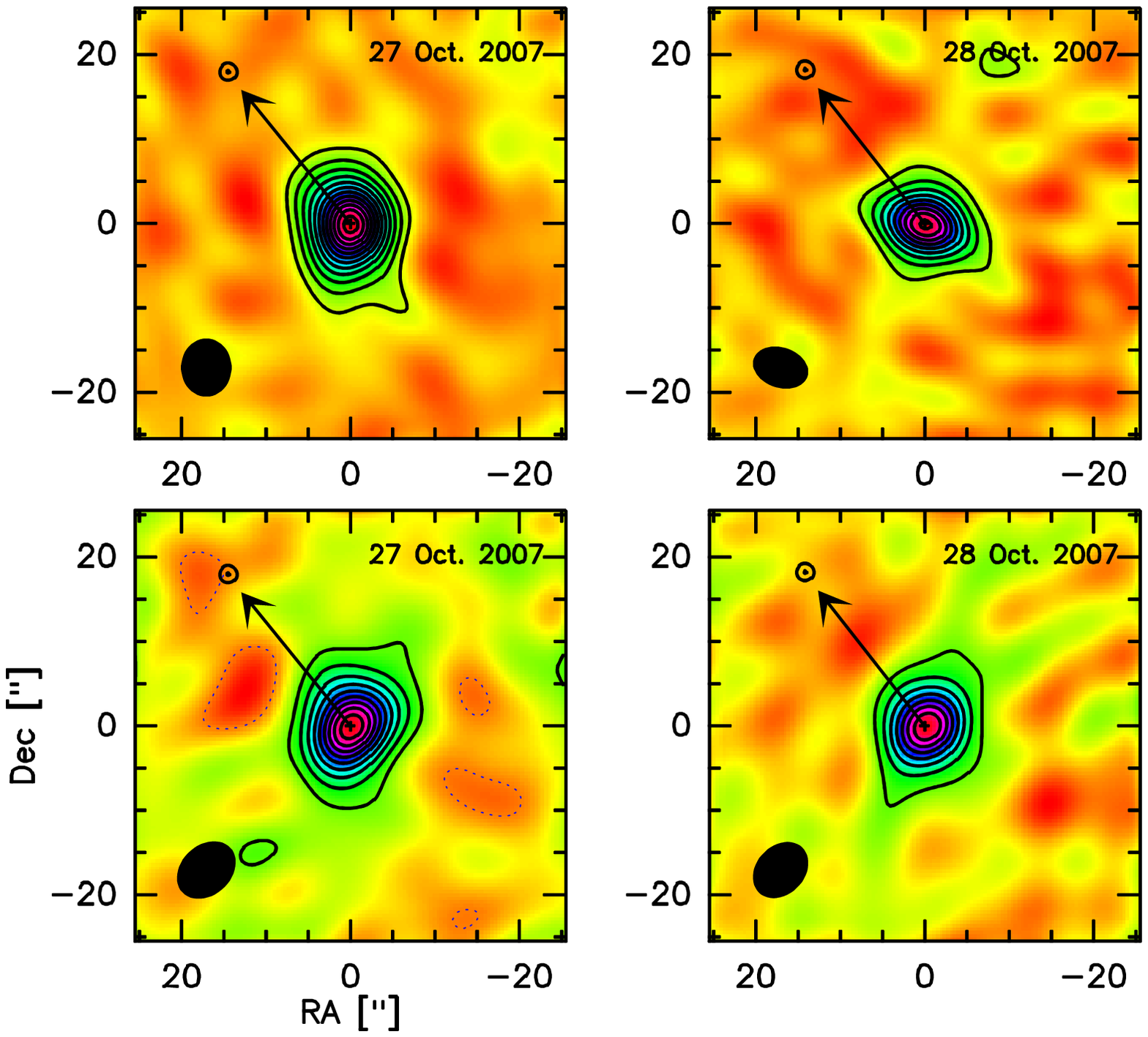}}
    \caption{Interferometric maps of the 3.3 mm continuum emission of comet 17P/Holmes obtained on
    27 and 28 October 2007 UT with PdBI. Top: full set of measurements. Bottom: maps constructed using the
    same $uv$-coverage, i.e., from data acquired with 5 antennas and between 2h00 and 7h00 UT
 for both dates. The interferometric beam is plotted in the bottom left
 corner. The maps have been centred on the position of maximum brightness.
 The arrow represents the Sun direction projected on the sky. The level spacing is 2$\sigma$, with
 $\sigma$ = 0.093 (top figure) and 0.10 (bottom) mJy/beam for 27 October, and $\sigma$ = 0.12 (top) and 0.14  (bottom) mJy/beam for  28 October.}
\label{fig-maps0}
\end{figure}

The interferometric maps deduced from the whole data set are
presented in Fig.~\ref{fig-maps0} and their main characteristics
(flux density at the centre of the map and astrometric position of
the peak of brightness) are reported in Tables~\ref{log} and
~\ref{pos}. Note that the maps have been centred on the position
of the peak of brightness. The comparison
of the 27 and 28 October maps is difficult due to differences in
the shape and size of the interferometric beam, caused by
different $uv$-coverages. Maps obtained with a similar
$uv$-coverage using restricted data sets (from 2h00 to 7h00 UT,
same 5 antennas for both dates) are also presented in
Fig.~\ref{fig-maps0}, and their characteristics are given in the
second part of Tables~\ref{log}--\ref{pos}.

\section{Data analysis}
\label{sec:2}

The detected 3.3 mm emission is due to the thermal emission of
dust particles in the coma. Indeed, a flux density of 2 mJy at 90
GHz would correspond to a photometric diameter of 60 km for a spherical
slow rotator at the equilibrium temperature of 207 K expected at
2.45 AU from the Sun. Instead, the effective diameter of the
nucleus of 17P/Holmes is estimated to be 3.2 km \citep{Snod2006}.

The first interesting feature in these 3.3-mm continuum
measurements is the small variation of the flux density in 24
hours elapsed time. The peak flux density within the $\sim$ 5--6$\arcsec$
field-of-view decreased by 20--25\% only between the two dates
(Table~\ref{log}).

\subsection{Astrometry and azimuthal variations}

As seen in Table~\ref{pos}, the astrometric position of the
maximum brightness (O) almost coincides with the position of the
nucleus given by the ephemeris (C). An (O-C) offset of $\sim$
+0.3$\arcsec$ at the 1--$\sigma$ confidence level is measured in Declination for
both dates when considering the whole data set. On optical images,
the inner coma of 17P/Holmes presents an asymmetric distribution,
mainly characterized by the presence at the position angle p.a. =
220$^\circ$ (corresponding to the anti-Sun direction) of a bright
elongated cloud (so-called ''blob''), quickly separating
at a rate of 9--10 $\arcsec$/day from the (much brighter) condensation
of material surrounding the nucleus
\citep{montalto2008,Wata2009,reach10}. The (O-C) marginal offsets
measured in the PdBI maps (p.a. within 0--60$^\circ$) are not towards the direction of
this expanding blob: besides the uncertainty of the measurements
and of the comet ephemeris, they can be due to spatial asymmetries
of the brightness distribution in either direction. At the time
of our observations, the blob was at a projected distance of
$\sim$ 17$\arcsec$ (27 October) and 27$\arcsec$ (28 October) from the nucleus
\citep{montalto2008,Wata2009}, i.e., at the edge of the primary
beam of the PdBI antennas that defines the extent of the
interferometric map. Its thermal emission is not seen in our
images (Fig.~\ref{fig-maps0}). Thus, the amount of
millimetre-sized dust particles in this blob was too low with
respect to the amount of material surrounding the nucleus for a
detection in the PdBI maps. We have also to consider that more
massive particles have lower expansion velocities and are less
subject to radiation pressure, so that this bright blob of
micrometric particles seen in optical and infrared images was
likely deficient in particles radiating at 3.3 mm. Comparing
optical and infrared images, \citet{reach10} concluded that the
blob comprised particles of intermediate ($\sim$ 10 $\mu$m) sizes.

\begin{figure}
\resizebox{\hsize}{!}{\includegraphics[angle=0]{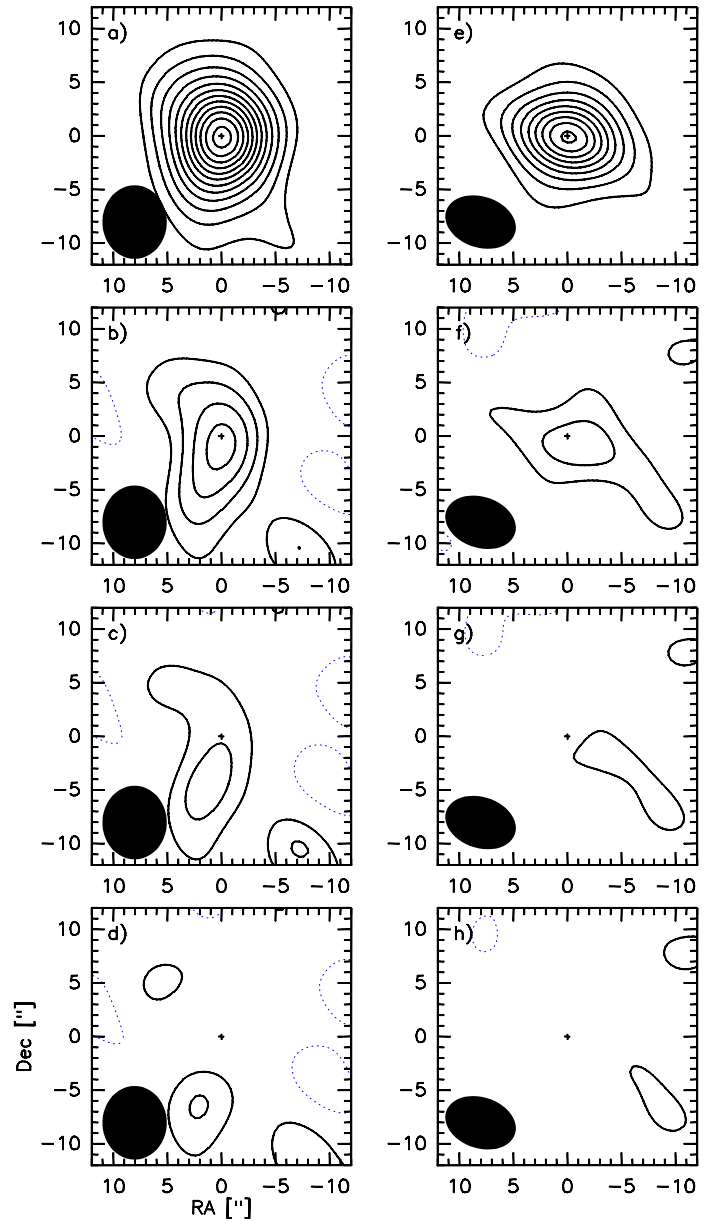}}
    \caption{Asymmetries in the interferometric 3.3 mm maps of comet 17P/Holmes from image processing. Results for 27 and 28 October are plotted in left {\bf (a--d)} and right {\bf (e--h)} panels, respectively.       
Top figures ({\bf a}, {\bf e}): observed images as in Fig.~ \ref{fig-maps0}. All other figures have had a symmetric image subtracted (in the Fourier plane) to enhance azimuthal asymmetries. The symmetric images for  27 and 28 October are described by the fits of the visibilities given in Fig.~\ref{visi}, multiplied by a factor $f$ equal to 0.8 ({\bf b}, {\bf f}), 0.9 ({\bf c}, {\bf g}), and 1.0 ({\bf d}, {\bf h}). The interferometric beam is plotted in the bottom left corner. The level spacing is 1$\sigma$, except for the top figures (2$\sigma$, see Fig.~\ref{fig-maps0}), with  $\sigma$ = 0.093 mJy/beam for images ({\bf a--d}), and $\sigma$ = 0.12 mJy/beam for ({\bf e--h}). Dashed contours correspond to negative fluxes. } 
\label{fig-map-diff}
\end{figure}

Spatial asymmetries are marginally present in the PdBI maps. To enhance spatial 
features, we subtracted a symmetric image to the measurements. The subtraction was done
in the Fourier plane, that is, we subtracted the visibilities characterizing the symmetric image to the measured visibilities. Interferometric maps were then produced with the new sets of visibilities (i.e., $uv$-tables). The visibility amplitude $\mathcal{V}$ of the symmetric images follows $\mathcal{V}$ = 0.052$r_{uv}^{-0.83}$ Jy and $\mathcal{V}$ = 0.025$r_{uv}^{-0.66}$ Jy, for  27 and 28 October, respectively. This corresponds to the 
angular average of the data in the Fourier plane described in Sect.~\ref{sec:radial}. Figures~\ref{fig-map-diff}d, h show, for the two observation dates, the {\it residual} interferometric images. We also plot residual images obtained by applying a factor $f$ = 0.8 (Fig.~\ref{fig-map-diff}b, f) and 0.9 (Figs.~\ref{fig-map-diff}c,~\ref{fig-map-diff}g) to visibilities of the symmetric images. An emission feature is observed at the 2$\sigma$ level  South-East from the nucleus position (p.a. $\sim$ 160$^\circ$), in the 27 October residual images obtained for $f$ = 0.9 and 1.0. As a matter of fact, this feature is discernable on the 27 October original map  (Figs.~\ref{fig-maps0} left and~\ref{fig-map-diff}a). Its intensity is $\sim$ 10\% the peak intensity in the centre of the 27 October map. This feature is not present in the 28 October images. More marginal features at the noise level (1$\sigma$) are observed  at position angles corresponding approximatively to the Sun direction (27 October), and to the tail direction (28 October)  (p.a.(Sun) = 38$^{\circ}$, p.a.(tail) = 218$^{\circ}$). All these features are also observed when the restricted data sets are used. We show in Sect.~\ref{sec:5} that 
the velocity of the 1-mm sized particles in the expanding shell is 50--100 m s$^{-1}$. 
This is compatible with the disappearance of the southward feature in the PdBI map of 28 October.

\subsection{Radial distribution}
\label{sec:radial}

The data provide information on the radial distribution of the
emission at projected distances $D$ of typically 3500 to 20\,000 km
from the nucleus (this corresponds to the range of $\lambda$/$B_l$
projected at the distance of the comet, where $B_l$ is the
baseline length and $\lambda$ is the wavelength). For comets
observed in steady-state activity, the column density and
brightness distribution of the dust in the inner coma is expected
to vary in first approximation according to 1/$D$, not considering
deviations from this ideal law caused by radiation pressure,
particle fragmentation, and asymmetries related to the nucleus
outgassing geometry. There are several examples of surface
brightness profiles at millimetre and submillimetre wavelengths
consistent with this ideal law \citep[e.g.,][]{jewmat99,dbm2010b}. On the
other hand, when observations are conducted soon after a massive outburst, the radial profile may strongly differ from this law. The profile may resemble that of a point-like
source, if the size of the expanding cloud is much smaller than the
angular resolution of the interferometric beam. Taking into
account the elapsed time between the outburst onset time and the
PdBI observations, this would happen for grains with velocities
significantly less than 10 m/s. Conversely, if the millimetre-sized
particles that contribute to the emission have sufficiently high
velocities, a ring-like structure, arcs or bright condensations
(not coinciding with the nucleus position) can be observed,
depending on the geometry of the expansion. In the present case, this is not
observed. Finally, a variety of profiles,
eventually approaching the 1/$D$ law in restricted regions of the
coma, can be observed if the outburst was followed by a sustained
production of millimetre-sized particles (e.g., due to dust
fragmentation) and/or if the particles contributing to the signal
have a broad range of velocities. The decline of the dust emission
from  27 and 28 October is governed both by the production curve of
the particles and their velocity.

\begin{figure}
\includegraphics[angle=-90, width = 10 cm]{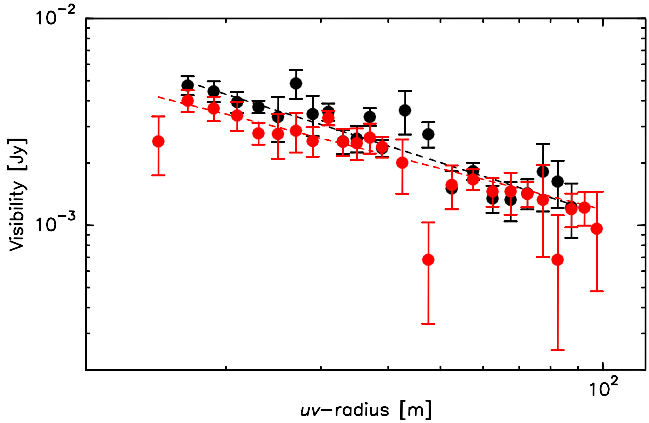}
\caption{Visibilities measured on 27  (black symbols) and
28 (red symbols) October. Dashed lines show fitted dependences with
$uv$-radius: 0.052$r_{uv}^{-0.83}$ (27 October) and
0.025$r_{uv}^{-0.66}$ (28 October) in janskys.} \label{visi}
\end{figure}

As discussed in \citet{boissier2007}, an appropriate way to study
the radial distribution of material observed by interferometric
techniques is from the analysis of the visibilities which are the
direct output of interferometers. This method allows us to avoid
uncertainties in the maps resulting from the partial $uv$-coverage (see Fig.~\ref{fig-uv}).
In addition, due to filtering and missing data at short
antenna-spacings, maps  lose information concerning the comet
emission at large scales (see Fig.~\ref{fig-maps0} where the
emission is barely detected at $D > 10\arcsec$). Observed visibilities
have been azimuthally averaged in the $uv$-plane to investigate the
radial brightness distribution. Figure~\ref{visi} 
displays the visibility amplitudes  
$\mathcal{V}$ as a function of $uv$-radius $r_{uv}$ (the baseline
length projected onto the plane on the sky) for the two dates. They vary according to $r_{uv}^{-0.83\pm0.06}$
and $r_{uv}^{-0.66\pm0.07}$ for  27 and 28 October, respectively.
The radial variation of the visibilities is indicative of a more
compact distribution than expected for a 1/$D$ variation of the
column density. For such a distribution, a $r_{uv}^{-1}$ variation
would have been observed \citep{dbm10} ($\mathcal{V}$ independent
of $uv$-radius characterizes compact sources unresolved by the
interferometic beam). A 23\% decline of the emission is observed
at $r_{uv} \leq$ 30 m between the two dates, whereas the visibility at
$r_{uv}$ $>$ 60 m is constant, within the uncertainties.

\section{Modelling}
\label{sec:4}

\begin{table}
\begin{center}
\caption{Optical constants $m = n - ik$ at 3.3 mm.} \label{indices}
\begin{tabular}{l c c c c}
\hline \noalign{\smallskip} \hline \noalign{\smallskip} Compound &
\phantom{0}$n$\phantom{0} & \phantom{0}$k$\phantom{0} & Reference \\
\hline\noalign{\smallskip}
Water ice & 1.79 & 2.00$\times$10$^{-3}$& Warren \& Brandt (2008) \\
Astro. silicates & 3.44 & 1.50$\times$10$^{-2}$ & Draine (1985) \\
Forsterite &2.63& 2.06$\times$10$^{-4}$ & Fabian et al. (2001) \\
Organics & 2.28 & 2.50$\times$10$^{-3}$ & Pollack et al. (1994)\\
Silicate mixt.$^a$ & 2.76 & 2.82$\times$10$^{-3}$ & this work\\
 \hline
\end{tabular}
\begin{list}{}{}
\item[$^a$] 20\% astronomical silicates and 80\% forsterite.
\end{list}
\end{center}
\end{table}

\begin{table*}
\begin{center}
\caption{Dust opacities at 3.3 mm.} \label{tab:opacity}
\begin{tabular}{l c c c c c c }
\hline \noalign{\smallskip} \hline \noalign{\smallskip}
Compound & Size index & Size range & Porosity  & & $\kappa_{\rm 3.3mm}$ (m$^2$ kg$^{-1}$) &\\
 &   &  (mm) &        & &  & \\
 \cline{5-7}\\
&&&&Ice fraction = 0 \%& Ice fraction = 22 \%& Ice fraction = 48 \%\\
\hline\noalign{\smallskip}
Astro. silicates & -3.0 & 10$^{-4}$--1\phantom{000}  & (0, 0.5, 0.8) & (8.14, 5.42, 4.09) 10$^{-2}$  & (3.72, 2.26, 2.22) 10$^{-2}$ & (1.95, 1.49, 1.53) 10$^{-2}$\\
 Astro. silicates & -3.0 & 10$^{-4}$--10\phantom{00}  & (0, 0.5, 0.8) & (3.99, 5.03, 6.16) 10$^{-2}$ & (3.07, 3.35, 3.30) 10$^{-2}$ & (2.34, 2.39, 2.09) 10$^{-2}$\\
 Astro. silicates & -3.0 & 10$^{-4}$--100\phantom{0}  & (0, 0.5, 0.8) & (0.88, 1.51, 2.81) 10$^{-2}$ & (0.92, 1.43, 2.15) 10$^{-2}$ & (0.94, 1.32, 1.67) 10$^{-2}$ \\
 Astro. silicates & -3.5 & 10$^{-4}$--1\phantom{000}  & (0, 0.5, 0.8) & (5.39, 3.65, 3.43) 10$^{-2}$ & (2.44, 1.75, 2.03) 10$^{-2}$ & (1.38, 1.25, 1.44) 10$^{-2}$  \\
Astro. silicates & -3.5 & 10$^{-4}$--10\phantom{00}  & (0, 0.5, 0.8) & (4.62, 4.98, 5.62) 10$^{-2}$& (3.12, 3.08, 2.98) 10$^{-2}$& (2.23, 2.15, 1.91) 10$^{-2}$ \\
Astro. silicates & -3.5 & 10$^{-4}$--100\phantom{0}  & (0, 0.5, 0.8) & (1.89, 2.48, 3.68) 10$^{-2}$& (1.49, 1.91, 2.43) 10$^{-2}$& (1.31, 1.58, 1.77) 10$^{-2}$\\
Astro. silicates & -3.5 & 10$^{-4}$--1000  & (0, 0.5, 0.8) & (0.64, 0.92, 1.46) 10$^{-2}$& (0.55, 0.75, 1.14) 10$^{-2}$& (0.53, 0.70, 0.98) 10$^{-2}$ \\
Astro. silicates & -4.0 & 10$^{-4}$--1\phantom{000}  & (0, 0.5, 0.8) & (1.78, 1.71, 2.68) 10$^{-2}$  & (0.98, 1.18, 1.79) 10$^{-2}$ & (0.73, 0.98, 1.33) 10$^{-2}$\\
Astro. silicates & -4.0 & 10$^{-4}$--10\phantom{00}  & (0, 0.5, 0.8)  & (2.48,
2.63, 3.50) 10$^{-2}$ & (1.58, 1.72, 2.11) 10$^{-2}$ & (1.16, 1.30, 1.48) 10$^{-2}$\\
Astro. silicates & -4.0 & 10$^{-4}$--100\phantom{0}  & (0, 0.5, 0.8)  &
(2.17, 2.46, 3.45) 10$^{-2}$ & (1.40, 1.69, 2.15) 10$^{-2}$ & (1.14, 1.33, 1.53) 10$^{-2}$ \\
 Silicate mixt. & -3.5 & 10$^{-4}$--1\phantom{000}  & (0, 0.5, 0.8) & (0.80, 0.52, 0.60) 10$^{-2}$ & (0.84, 0.68, 0.79) 10$^{-2}$ & (0.87, 0.85, 0.96) 10$^{-2}$ \\
 Silicate mixt. & -3.5 & 10$^{-4}$--10\phantom{00}  & (0, 0.5, 0.8) & (1.13, 1.00, 0.93) 10$^{-2}$ & (1.07, 1.02, 0.88) 10$^{-2}$ & (1.16, 1.04, 0.91) 10$^{-2}$ \\
Silicate mixt. & -3.5 & 10$^{-4}$--100\phantom{0}  & (0, 0.5, 0.8) & (0.61, 0.72, 0.83) 10$^{-2}$ & (0.68, 0.77, 0.85) 10$^{-2}$ & (0.82, 0.88, 0.89) 10$^{-2}$ \\
Organics & -3.5 & 10$^{-4}$--100\phantom{0}  & (0, 0.5, 0.8) & (0.89, 0.97, 1.05) 10$^{-2}$  & (0.87, 0.96, 0.99) 10$^{-2}$  & (0.90, 0.97, 0.99) 10$^{-2}$  \\
\hline
\end{tabular}
\end{center}
\end{table*}

\onltab{5}{
\begin{table*}
\begin{center}
\caption{Dust opacities at 0.45, 0.8, and 1.2 mm.} \label{tab:opacity-otherfreq}
\begin{tabular}{l c c c c c c }
\hline \noalign{\smallskip} \hline \noalign{\smallskip}
Compound & Size index & Size range & Porosity  & & $\kappa_{\rm \lambda}$ (m$^2$ kg$^{-1}$) &\\
 &   &  (mm) &        & &  & \\
 \cline{5-7}\\
&&&&Ice fraction = 0 \%& Ice fraction = 22 \%& Ice fraction = 48 \%\\
\hline\noalign{\smallskip}
\underline{$\lambda$ = 1.2 mm } &&&&&&\\
 Astro. silicates & -3.0 & 10$^{-4}$--1\phantom{000}  & (0, 0.5, 0.8) & (3.99, 4.67, 4.51) 10$^{-1}$  & (3.00, 2.72, 2.23) 10$^{-1}$ & (2.17, 1.70, 1.43) 10$^{-1}$\\
 Astro. silicates & -3.0 & 10$^{-4}$--10\phantom{00}  & (0, 0.5, 0.8) & (0.93, 1.58, 2.70) 10$^{-1}$ & (0.96, 1.40, 1.96) 10$^{-1}$ & (0.96, 1.23, 1.43) 10$^{-1}$\\
 Astro. silicates & -3.0 & 10$^{-4}$--100\phantom{0}  & (0, 0.5, 0.8) & (1.44, 2.73, 5.86) 10$^{-2}$ & (1.70, 3.04, 6.05) 10$^{-2}$ & (1.96, 3.32, 6.03) 10$^{-2}$ \\
 Astro. silicates & -3.5 & 10$^{-4}$--1\phantom{000}  & (0, 0.5, 0.8) & (3.48, 3.72, 3.70) 10$^{-1}$ & (2.39, 2.13, 1.95) 10$^{-1}$ & (1.68, 1.39, 1.29) 10$^{-1}$  \\
 Astro. silicates & -3.5 & 10$^{-4}$--10\phantom{00}  & (0, 0.5, 0.8) & (1.59, 2.19, 3.11) 10$^{-1}$& (1.33, 1.64, 2.03) 10$^{-1}$& (1.18, 1.33, 1.42) 10$^{-1}$ \\
Astro. silicates & -3.5 & 10$^{-4}$--100\phantom{0}  & (0, 0.5, 0.8) & (0.56, 0.80, 1.28) 10$^{-1}$& (0.49, 0.67, 1.01) 10$^{-1}$& (0.47, 0.61, 0.85) 10$^{-1}$\\
 Astro. silicates & -4.0 & 10$^{-4}$--1\phantom{000}  & (0, 0.5, 0.8)  & (1.56,
1.80, 2.45) 10$^{-1}$ & (1.07, 1.19, 1.53) 10$^{-1}$ & (0.81, 0.90, 1.09) 10$^{-1}$\\
 Astro. silicates & -4.0 & 10$^{-4}$--10\phantom{00}  & (0, 0.5, 0.8)  & (1.42,
1.81, 2.61) 10$^{-1}$ & (1.05, 1.28, 1.66) 10$^{-1}$ & (0.87, 1.01, 1.18) 10$^{-1}$\\
 Astro. silicates & -4.0 & 10$^{-4}$--100\phantom{0}  & (0, 0.5, 0.8)  &
(1.22, 1.53, 2.26) 10$^{-1}$ & (0.90, 1.11, 1.49) 10$^{-1}$ & (0.75, 0.90, 1.10) 10$^{-1}$\\
 Silicate mixt. & -3.0 & 10$^{-4}$--1\phantom{000}  & (0, 0.5, 0.8) & (1.07, 0.92, 0.69) 10$^{-1}$  & (1.05, 0.81, 0.66) 10$^{-1}$ & (1.10, 0.81, 0.67) 10$^{-1}$\\
 Silicate mixt. & -3.0 & 10$^{-4}$--10\phantom{00}  & (0, 0.5, 0.8) & (0.42, 0.56, 0.69) 10$^{-1}$ & (0.51, 0.61, 0.70) 10$^{-1}$ & (0.62, 0.70, 0.73) 10$^{-1}$\\
 Silicate mixt. & -3.0 & 10$^{-4}$--100\phantom{0}  & (0, 0.5, 0.8) & (0.98, 1.66, 3.05) 10$^{-2}$ & (1.25, 2.08, 3.59) 10$^{-2}$ & (1.58, 2.56, 4.12) 10$^{-2}$ \\
 Silicate mixt. & -3.5 & 10$^{-4}$--1\phantom{000}  & (0, 0.5, 0.8) & (0.84, 0.70, 0.59) 10$^{-1}$ & (0.80, 0.64, 0.59) 10$^{-1}$ & (0.84, 0.67, 0.62) 10$^{-1}$  \\
 Silicate mixt. & -3.5 & 10$^{-4}$--10\phantom{00}  & (0, 0.5, 0.8) & (0.53, 0.61, 0.68) 10$^{-1}$& (0.60, 0.64, 0.68) 10$^{-1}$& (0.70, 0.72, 0.70) 10$^{-1}$ \\
 Silicate mixt. & -3.5 & 10$^{-4}$--100\phantom{0}  & (0, 0.5, 0.8) & (0.24, 0.30, 0.42) 10$^{-1}$& (0.26, 0.34, 0.46) 10$^{-1}$& (0.30, 0.39, 0.51) 10$^{-1}$\\
Silicate mixt. & -4.0 & 10$^{-4}$--1\phantom{000}  & (0, 0.5, 0.8)  & (0.35,
0.36, 0.45) 10$^{-1}$ & (0.37, 0.40, 0.49) 10$^{-1}$ & (0.42, 0.46, 0.54) 10$^{-1}$\\
 Silicate mixt. & -4.0 & 10$^{-4}$--10\phantom{00}  & (0, 0.5, 0.8)  & (0.37,
0.41, 0.51) 10$^{-1}$ & (0.41, 0.45, 0.54) 10$^{-1}$ & (0.48, 0.52, 0.58) 10$^{-1}$\\
 Silicate mixt. & -4.0 & 10$^{-4}$--100\phantom{0}  & (0, 0.5, 0.8)  &
(0.36, 0.38, 0.48) 10$^{-1}$ & (0.36, 0.42, 0.51) 10$^{-1}$ & (0.42, 0.48, 0.56) 10$^{-1}$\\
&&&&&&\\
\underline{$\lambda$ = 0.8 mm } &&&&&&\\
 Astro. silicates & -3.0 & 10$^{-4}$--1\phantom{000}  & (0, 0.5, 0.8) & (6.04, 8.17, 9.90) 10$^{-1}$ & (5.09, 5.70, 5.17) 10$^{-1}$ & (4.19, 3.94, 3.33) 10$^{-1}$ \\
 Astro. silicates & -3.0 & 10$^{-4}$--10\phantom{00}  & (0, 0.5, 0.8) & (1.16, 2.06, 3.93) 10$^{-1}$ & (1.28, 2.05, 3.31) 10$^{-1}$ & (1.37, 1.99, 2.70) 10$^{-1}$ \\
 Astro. silicates & -3.0 & 10$^{-4}$--100\phantom{0}  & (0, 0.5, 0.8) & (1.64, 3.20, 7.08) 10$^{-2}$ & (2.02, 3.68, 7.68) 10$^{-2}$ & (2.38, 4.16, 8.13) 10$^{-2}$ \\
 Astro. silicates & -3.5 & 10$^{-4}$--1\phantom{000}  & (0, 0.5, 0.8) & (6.02, 7.25, 8.46) 10$^{-1}$ & (4.55, 4.77, 4.57) 10$^{-1}$ & (3.56, 3.30, 3.03) 10$^{-1}$ \\
 Astro. silicates & -3.5 & 10$^{-4}$--10\phantom{00}  & (0, 0.5, 0.8) & (2.41, 3.47, 5.32) 10$^{-1}$ & (2.14, 2.83, 3.81) 10$^{-1}$ & (2.00, 2.42, 2.87) 10$^{-1}$ \\
 Astro. silicates & -3.5 & 10$^{-4}$--100\phantom{0}  & (0, 0.5, 0.8) & (0.81, 1.20, 1.97) 10$^{-1}$ & (0.76, 1.04, 1.61) 10$^{-1}$ & (0.72, 0.96, 1.41) 10$^{-1}$ \\
 Astro. silicates & -4.0 & 10$^{-4}$--1\phantom{000}  & (0, 0.5, 0.8) & (3.19, 3.97, 5.68) 10$^{-1}$ & (2.34, 2.79, 3.52) 10$^{-1}$ & (1.89, 2.13, 2.51) 10$^{-1}$ \\
 Astro. silicates & -4.0 & 10$^{-4}$--10\phantom{00}  & (0, 0.5, 0.8) & (2.74, 3.62, 5.46) 10$^{-1}$ & (2.12, 2.69, 3.59) 10$^{-1}$ & (1.82, 2.17, 2.61) 10$^{-1}$ \\
 Astro. silicates & -4.0 & 10$^{-4}$--100\phantom{0}  & (0, 0.5, 0.8) & (2.29, 3.03, 4.63) 10$^{-1}$ & (1.81, 2.28, 3.11) 10$^{-1}$ & (1.54, 1.86, 2.32) 10$^{-1}$ \\

 Silicate Mixt. & -3.0 & 10$^{-4}$--1\phantom{000}  & (0, 0.5, 0.8) & (1.97, 1.99, 1.64) 10$^{-1}$ & (2.06, 1.98, 1.56) 10$^{-1}$ & (2.42, 1.95, 1.61) 10$^{-1}$ \\
 Silicate Mixt. & -3.0 & 10$^{-4}$--10\phantom{00}  & (0, 0.5, 0.8) & (0.63, 0.93, 1.29) 10$^{-1}$ & (0.80, 1.07, 1.38) 10$^{-1}$ & (0.98, 1.27, 1.50) 10$^{-1}$ \\
 Silicate Mixt. & -3.0 & 10$^{-4}$--100\phantom{0}  & (0, 0.5, 0.8) & (1.17, 2.12, 4.22) 10$^{-2}$ & (1.56, 2.73, 5.17) 10$^{-2}$ & (1.99, 3.40, 6.18) 10$^{-2}$ \\
 Silicate Mixt. & -3.5 & 10$^{-4}$--1\phantom{000}  & (0, 0.5, 0.8) & (1.72, 1.62, 1.42) 10$^{-1}$ & (1.74, 1.60, 1.40) 10$^{-1}$ & (1.98, 1.63, 1.47) 10$^{-1}$ \\
 Silicate Mixt. & -3.5 & 10$^{-4}$--10\phantom{00}  & (0, 0.5, 0.8) & (0.90, 1.12, 1.36) 10$^{-1}$ & (1.07, 1.23, 1.42) 10$^{-1}$ & (1.26, 1.41, 1.52) 10$^{-1}$ \\
 Silicate Mixt. & -3.5 & 10$^{-4}$--100\phantom{0}  & (0, 0.5, 0.8) & (0.35, 0.47, 0.70) 10$^{-1}$ & (0.41, 0.55, 0.78) 10$^{-1}$ & (0.49, 0.64, 0.89) 10$^{-1}$ \\
 Silicate Mixt. & -4.0 & 10$^{-4}$--1\phantom{000}  & (0, 0.5, 0.8) & (0.81, 0.86, 1.03) 10$^{-1}$ & (0.87, 0.96, 1.13) 10$^{-1}$ & (1.03, 1.09, 1.24) 10$^{-1}$ \\
 Silicate Mixt. & -4.0 & 10$^{-4}$--10\phantom{00}  & (0, 0.5, 0.8) & (0.76, 0.89, 1.11) 10$^{-1}$ & (0.88, 1.00, 1.20) 10$^{-1}$ & (1.03, 1.16, 1.31) 10$^{-1}$ \\
 Silicate Mixt. & -4.0 & 10$^{-4}$--100\phantom{0}  & (0, 0.5, 0.8) & (0.68, 0.78, 1.01) 10$^{-1}$ & (0.75, 0.89, 1.09) 10$^{-1}$ & (0.88, 1.03, 1.21) 10$^{-1}$ \\

&&&&&&\\
\underline{$\lambda$ = 0.45 mm } &&&&&&\\
 Astro. silicates & -3.0 & 10$^{-4}$--1\phantom{000}  & (0, 0.5, 0.8) & (0.93, 1.50, 2.33) \phantom{10$^{00}$} & (0.94, 1.27, 1.53) \phantom{10$^{00}$} & (0.90, 1.06, 1.05) \phantom{10$^{00}$} \\
 Astro. silicates & -3.0 & 10$^{-4}$--10\phantom{00}  & (0, 0.5, 0.8) & (1.46, 2.74, 5.71) 10$^{-1}$ & (1.74, 3.00, 5.67) 10$^{-1}$ & (1.98, 3.23, 5.45) 10$^{-1}$ \\
 Astro. silicates & -3.0 & 10$^{-4}$--100\phantom{0}  & (0, 0.5, 0.8) & (1.95, 3.86, 8.85) 10$^{-2}$ & (2.45, 4.62, 9.99) 10$^{-2}$ & (2.96, 5.40, 11.2) 10$^{-2}$ \\
 Astro. silicates & -3.5 & 10$^{-4}$--1\phantom{000}  & (0, 0.5, 0.8) & (1.15, 1.61, 2.27) \phantom{10$^{00}$} & (1.02, 1.24, 1.43) \phantom{10$^{00}$} & (0.90, 0.99, 0.99) \phantom{10$^{00}$} \\
 Astro. silicates & -3.5 & 10$^{-4}$--10\phantom{00}  & (0, 0.5, 0.8) & (0.41, 0.63, 1.03) \phantom{10$^{00}$} & (0.40, 0.55, 0.83) \phantom{10$^{00}$} & (0.39, 0.51, 0.69) \phantom{10$^{-1}$} \\
 Astro. silicates & -3.5 & 10$^{-4}$--100\phantom{0}  & (0, 0.5, 0.8) & (1.35, 2.08, 3.56) 10$^{-1}$ & (1.31, 1.89, 3.01) 10$^{-1}$ & (1.30, 1.81, 2.72) 10$^{-1}$ \\
 Astro. silicates & -4.0 & 10$^{-4}$--1\phantom{000}  & (0, 0.5, 0.8) & (0.81, 1.12, 1.74) \phantom{10$^{00}$} & (0.67, 0.86, 1.14) \phantom{10$^{00}$} & (0.58, 0.70, 0.82) \phantom{10$^{00}$} \\
 Astro. silicates & -4.0 & 10$^{-4}$--10\phantom{00}  & (0, 0.5, 0.8) & (0.66, 0.94, 1.51) \phantom{10$^{00}$} & (0.56, 0.75, 1.04) \phantom{10$^{00}$} & (0.50, 0.63, 0.79) \phantom{10$^{00}$} \\
 Astro. silicates & -4.0 & 10$^{-4}$--100\phantom{0}  & (0, 0.5, 0.8) & (0.56, 0.79, 1.27) \phantom{10$^{00}$} & (0.47, 0.63, 0.88) \phantom{10$^{00}$} & (0.42, 0.53, 0.67) \phantom{10$^{00}$} \\
 Silicate mixt.  & -3.0 & 10$^{-4}$--1\phantom{000}  & (0, 0.5, 0.8) & (0.41, 0.49, 0.52) \phantom{10$^{00}$} & (0.49, 0.54, 0.51) \phantom{10$^{00}$} & (0.57, 0.60, 0.54) \phantom{10$^{00}$} \\
Silicate mixt. & -3.0 & 10$^{-4}$--10\phantom{00}  & (0, 0.5, 0.8) & (0.95, 1.58, 2.69) 10$^{-1}$ & (1.26, 1.97, 3.14) 10$^{-1}$ & (1.59, 2.42, 3.61) 10$^{-1}$ \\
Silicate mixt. & -3.0 & 10$^{-4}$--100\phantom{0}  & (0, 0.5, 0.8) & (1.48, 2.77, 5.91) 10$^{-2}$ & (2.02, 3.64, 7.52) 10$^{-2}$ & (2.60, 4.62, 9.25) 10$^{-2}$ \\
 Silicate mixt. & -3.5 & 10$^{-4}$--1\phantom{000}  & (0, 0.5, 0.8) & (0.41, 0.45, 0.46) \phantom{10$^{00}$} & (0.48, 0.48, 0.47) \phantom{10$^{00}$} & (0.54, 0.54, 0.50) \phantom{10$^{00}$} \\
 Silicate mixt. & -3.5 & 10$^{-4}$--10\phantom{00}  & (0, 0.5, 0.8) & (0.18, 0.24, 0.33) \phantom{10$^{00}$} & (0.22, 0.28, 0.37) \phantom{10$^{00}$} & (0.27, 0.33, 0.41) \phantom{10$^{-1}$} \\
Silicate mixt. & -3.5 & 10$^{-4}$--100\phantom{0}  & (0, 0.5, 0.8) & (0.63, 0.87, 1.35) 10$^{-1}$ & (0.77, 1.04, 1.57) 10$^{-1}$ & (0.93, 1.25, 1.82) 10$^{-1}$ \\
 Silicate mixt. & -4.0 & 10$^{-4}$--1\phantom{000}  & (0, 0.5, 0.8) & (2.41, 2.73, 3.38) 10$^{-1}$ & (2.81, 3.18, 3.74) 10$^{-1}$ & (3.31, 3.73, 4.18) 10$^{-1}$ \\
 Silicate mixt. & -4.0 & 10$^{-4}$--10\phantom{00}  & (0, 0.5, 0.8) & (2.09, 2.53, 3.31) 10$^{-1}$ & (2.50, 2.98, 3.70) 10$^{-1}$ & (3.00, 3.52, 4.15) 10$^{-1}$ \\
Silicate mixt. & -4.0 & 10$^{-4}$--100\phantom{0}  & (0, 0.5, 0.8) & (1.79, 2.14, 2.85) 10$^{-1}$ & (2.12, 2.54, 3.21) 10$^{-1}$ & (2.52, 3.00, 3.62) 10$^{-1}$ \\

\hline
\end{tabular}
\end{center}
\end{table*}
}

\subsection{Dust thermal emission}
\label{sec:thermal}

Models of the thermal emission of cometary dust in the microwave
domain have been presented by \citet{jew90,jew92}. The radiation
of dust grains at $\lambda$ = 3.3 mm depends of the grain
absorption efficiency factor $Q_{\rm abs}$, which is a function of the
grain size and of the complex refractive index $m_\lambda$ of the
material. We calculated $Q_{\rm abs}$ at $\lambda$ = 3.3 mm for
different mixtures of silicates, organics and water ice using the Mie theory
for spherical and homogenous grains. Infrared signatures of water
ice grains were observed in the coma of 17P/Holmes, and lifetime
calculations indicate that dirty and pure ice grains with $a$ $>$100
$\mu$m survived for several days or more after the outburst
\citep{Yang09}. We also considered the porosity of the grains. The
effective $m_\lambda$ of fluffy grains was computed in a two-step
process. First, we computed the refractive index of basic units
made of silicates and water ice using the Maxwell Garnett
effective medium theory (EMT) following \citet{green90}.
Mixtures of organics and ice, as well as mixtures of different
silicates were also considered (Table~\ref{indices}). The effective refractive
index of the fluffy particles was then deduced using the Maxwell
Garnett formula for two-component mixtures, taking vacuum for the
core material and the silicate/ice or organic/ice mixture in the
matrix (hollow sphere). This prescription was used by
\citet{kru94} for modelling $Q_{\rm abs}$ of interstellar dust. EMT
theories are only an approximation to the real optical behavior of
composite media, and give more accurate results when the size of
the grains is small relative to the wavelength, and when the
volume of the inclusions (here silicates, organics and voids) is
small with respect to the matrix \citep{Ossen91,Perrin90}. The use
of more exact theories is beyond the scope of this paper.

The refractive indices used in this study are given in Table~\ref{indices}. We
considered different materials, with optical constants which might
be representative of cometary material \citep{hanner}. Optical constants at 3.3
mm are rare in the literature. The value for the silicate mixture
made of forsterite and astronomical silicates was computed with
the Maxwell Garnett formula, with forsterite constituting the
matrix. We considered aggregates with a relative ice mass
fraction $M_{ice}$/($M_{ice}$+$M_{dirt}$) from 0 to 48 \% (where
dirt refers to silicates or organics) and a porosity from 0 to
0.8.

\begin{figure}
\includegraphics[angle=-90, width = 9.5 cm]{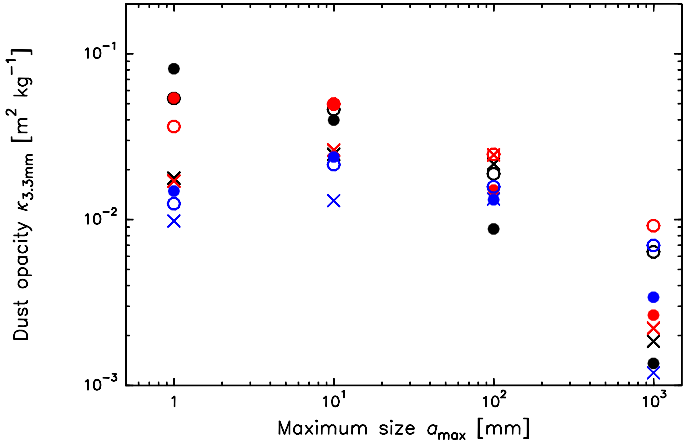}
\caption{ Dust opacities at 3.3 mm for grains made of astronomical silicates. Values for 
0\% ice and porosity = 0, 0\% ice and porosity = 50\%, 48\% ice and porosity = 50\% are plotted with black, red, and blue symbols, respectively. Plain, open circles, and crosses are for size indexes $q$ = --3, --3.5, and --4, respectively.} \label{fig: dust-opacities}
\end{figure}

\begin{figure}[t]
\resizebox{10.5cm}{!}{\includegraphics[angle=-90,bb = 110 17 530 674]{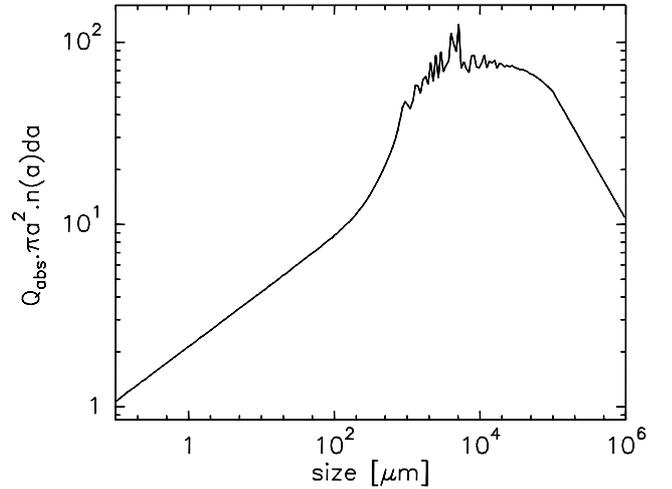}}
\caption{Product $Q_{\rm abs}(a)\times\pi
a^2\times n(a)da$ for a size distribution in $a^{-3.7}$ and absorption efficiencies at 3.3 mm of astronomical silicates with 0.5 porosity and 48\% ice fraction.}
\label{fig:para}
\end{figure}

If the dust properties (size distribution and chemical properties)
do not vary within the instrumental field of view, the flux
density (W m$^{-2}$ Hz$^{-1}$ or Jy) may be written as (e.g.,
Jewitt \& Luu 1990) :

\begin{equation}
S_\lambda = \frac{2 k }{\lambda^2 \Delta^2} \int_{a_{min}}^{a_{max}} T(a) Q_{\rm abs} \pi a^2n(a)da,
\end{equation}

\noindent where $n(a)$ $\propto$ $a^{q}$ is the size distribution,
$q$ is the size index in the coma, and $a_{min}$ and $a_{max}$ are the minimum
and maximum grain radii. $T(a)$ is the temperature of the grains
which, for the purpose of modelling thermal emission at 3.3 mm,
can be assumed in first approximation to be independent of size \citep{jew90}
and described by the blackbody equilibrium temperature of fast-rotating bodies of 174 K at
2.45 AU. The Rayleigh-Jeans approximation for the blackbody grain emission applies. The flux density is then related to the mass of emitting
grains $M$ through the so-called dust opacity $\kappa_\lambda$:
\begin{equation}
\label{eq:opa}
 S_\lambda = \frac{2 k T M \kappa_\lambda}{\lambda^2
\Delta^2},
\end{equation}

\noindent
with :

\begin{equation}
\kappa_\lambda = \frac{ \int_{a_{min}}^{a_{max}} Q_{\rm abs} \pi
a^2n(a)da }{ \int_{a_{min}}^{a_{max}}(4\pi /3) \rho a^3n(a)da },
\end{equation}

\noindent where $\rho$ is the density of the grains. Densities of
1000, 1500, 2500 kg m$^{-3}$ were taken for ice, organics, and
silicates, respectively. To compute $\kappa_\lambda$, we used the effective density of the grains, which depends
on the relative proportions of the materials and their porosity. For example, $\rho$ = 723 kg m$^{-3}$ for  silicate grains with 50\% ice content and a porosity of 0.5 (nominal composition considered in Sects. 4.2 and 4.3).
Sample opacities at $\lambda$ = 1 mm for non-porous grains made of
various (one component) materials have been presented by
\citet{jew92}. Table~\ref{tab:opacity} presents dust opacities at 3.3 mm
($k_{\rm 3.3 mm}$) for different ice/silicate mixtures, porosities
and size distributions.  The maximum size considered in the Table
is $a_{max}$ = 1000 mm, as evidence was found for 
cm-sized particles and even large-size fragments in 17P coma \citep{reach10,steven10}. When ice is included in the aggregates,
the opacity ranges from 0.005 and 0.037 m$^2$ kg$^{-1}$ in
the range of considered size indexes. For $a_{max}$ $>$ $\lambda$ = 3.3 mm, 
$\kappa_{\rm 3.3 mm}$ decreases with increasing $a_{max}$ because the larger particles, which are the most
efficient radiators at millimetre wavelengths, contribute to the
mass faster than they contribute to the radiating cross-section  (Fig.~\ref{fig: dust-opacities}).
On the other hand, $\kappa_{\rm 3.3 mm}$ shows little dependence with $a_{min}$.
The absorption efficiency shows resonances at particle radii
between 1 and 10 mm (i.e., at $\sim\lambda$, Fig.~\ref{fig:para}), which are smeared
out when the porosity increases. When the size distribution
encompasses these resonances ($a_{max}$ = 100 and 1000 mm in the
Table), porosity enhances the dust opacity at 3.3 mm when the
"hollow sphere" approximation is used  (Fig.~\ref{fig: dust-opacities}), whereas it does not
influence the dust opacity when vacuum is taken to be the matrix
in the Maxwell Garnett theory. For grains made of
astronomical silicates, $\kappa_{\rm 3.3 mm}$ decreases for
increasing amount of ice, as ice is more transparent than these
silicates. $\kappa_{\rm 3.3 mm}$ is also smaller for grains
containing organics or crystalline silicates (i.e., forsterite).

Applying Eq.~\ref{eq:opa} to the flux density of $\sim$ 2.4 mJy/beam  
observed on 27 October, we derive a dust mass $M$ = 0.8--6 $\times$ 10$^{11}$ kg for dust
opacities in the range 0.005--0.04 m$^2$ kg$^{-1}$. Obviously, this mass is a lower limit to the dust mass produced by the outburst, since the beam encompasses only part of the coma.

Using the same approach, we computed opacities at 0.45, 0.8 and 1.2 mm which might be useful for the interpretation of comet observations carried out with the Atacama Large Millimeter and submillimeter Array (ALMA) ($online$ Table~\ref{tab:opacity-otherfreq}).

\begin{table*}[t]
\begin{center}
\caption{One-component model: best fit for astronomical silicates with 0.5
porosity and 48\% ice fraction. } \label{result}
\begin{tabular}{l l c c c c c c  c c }
\hline \noalign{\smallskip} \hline \noalign{\smallskip}
Onset & Size  & $a_{max}$ & $V_0$ & $\chi^2_\nu$$^{(a)}$  & $\mathcal{V}_{\rm 28Oct}/\mathcal{V}_{\rm 27Oct}$ & \large{$\frac{F_{\rm 28Oct}}{F_{\rm 27Oct}}$} & $M_{\rm dust}^{\rm beam}$$^{(b, c)}$ & $M_{\rm dust}^{\rm tot}$$^{(c, d)}$  & $C_{\rm dust}^{\rm tot}$$^{(e)}$\\
(Oct. UT) & index & (mm) & (m s$^{-1}$) &  & $r_{uv }$ = 20 m& & (10$^{11}$ kg) & (10$^{11}$ kg) & (10$^{12}$ m$^2$) \\
\hline\noalign{\smallskip}
23.3    & $-$3 & 100 & 75 & 3.2 & 0.91 & 0.63 & 3.7 & 5.8 & 0.070 -- 0.083 \\
            & $-$3.5 & 100 & 60 & 1.9 & 0.93 & 0.73 & 3.0 & 5.4 &  1.8 -- 5.8 \\
            & $-$4 & 100 & 40 & 2.4 & 0.84 & 0.62 & 2.1 & 9.1 &  81 -- 802 \\
23.8  & $-$3 & 100 & 85 & 3.4 & 0.90 & 0.61 & 3.7  & 5.7 & 0.068 -- 0.082 \\
            & $-$3.5 & 100 & 65 & 2.3 & 0.88 & 0.64 & 3.1  & 5.2 &  1.7  -- 5.5\\
            & $-$4 & 100 & 35 & 1.7 & 0.87 & 0.65  & 2.3  & 7.9  & 71 -- 708 \\
23.8  & $-$3 & 1000 & 200 & 2.3 & 0.89 & 0.63  & 20 & 27 &  0.039 -- 0.045 \\
     & $-$3.5 & 1000 & 125 & 1.6 & 0.86 & 0.64 & 11  & 15 & 1.6 -- 5.2\\
     & $-$4 & 1000 & 40 & 1.5 & 0.85 & 0.66 & 3.5 & 9.4 &  71 -- 696\\
\hline\noalign{\smallskip}
Data &&&& & 0.84 $\pm$ 0.11 & 0.80 $\pm$ 0.05 &&& 55$^{(f)}$ \\
\hline
\end{tabular}
\end{center}
$^{(a)}$ Reduced $\chi^2$ from the fit of 27 and 28 October data.\\ 
$^{(b)}$ Dust mass within the synthesized beam on 27 October. \\
$^{(c)}$ Models with $a_{min}$ = 1 $\mu$m. Values obtained with $a_{min}$ = 0.1 $\mu$m differ by less than 20\%.\\
$^{(d)}$ Total dust mass produced by the outburst.\\
$^{(e)}$ Total scattering cross-section. Calculations are for $a_{min}$ in the range 0.1--1 $\mu$m. \\
$^{(f)}$ From optical observations \citep{Li2011}.
\end{table*}

\subsection{Time-dependent modelling}
\label{sec:model}
The in-depth analysis of the data requires a time-dependent model
of the dust coma. Indeed, since the outburst produced dust
particles and chunks with a broad range of velocities, the dust
size distribution evolved in the coma, both with time and distance
to nucleus. The velocity partitioning resulted in enhancing the
relative fraction of large particles with respect to smaller ones
in the inner coma, and might be observable on the visibility
curves which sample a broad range of distances to nucleus.
Alternatively, the visibilities at the different baseline lengths
(i.e., $uv$-radius) may sample particles produced with similar
kinematic properties but at different times.

The number of particles injected in the coma by the outburst as a
function of time is modelled by a function $G$($t$)
peaking at $t$ = $t_0$ \citep[23.3 October UT, according to][]{Hsieh2010}. $G$($t$) is described by the combination of two half Gaussians peaking at equal values, with widths $\Delta T_{\rm outburst}$ for $t > t_0$, and  $\Delta T_{\rm outburst}$/4 for $t < t_0$. The effective width of $G$($t$) is thus 5/8 $\times$ $\Delta T_{\rm outburst}$. This choice of function was initially motivated to accelerate the convergency of the computations. 

The size distribution of the dust injected by the outburst follows a power
law $n(a) \propto a^{q}$, defined by the size index $q$. The velocity of the dust particles is supposed to vary
according to :

\begin{equation}
\label{equ:4} V_{\rm dust}(a) = V_0~(a/a_0)^{-\delta},
\end{equation}

\noindent
where $V_0$ is the velocity of a grain of radius $a_0$ = 1 mm. We imposed $V_{\rm dust}$($a$) $<$ 1 km s$^{-1}$. It is worth mentioning that, for steady state dust production, the dust size index in the coma (that used in Sect.~\ref{sec:thermal} for computing dust opacities) is equal to $q+\delta$. 

We assumed a spherical dust coma. The density distribution of dust
grains of size $a$ is given by :
\begin{equation}
\label{eq:dust-density}
n_{\rm dust}(a,r) = \frac{n(a)}{4 \pi V_{\rm dust}(a) r^2} G (t-t_0-\frac{r}{V_{\rm dust}(a)}),
\end{equation}

\noindent where $r$ is the distance to the nucleus and $t$ the
time of the observation. The dust thermal emission by unit of
solid angle [Jy sr$^{-1}$] at impact parameter $D$ from the
nucleus is computed  by integration along the line of sight (LOS) according to:

\begin{equation}
\label{equ:6} 
F_{\lambda}(D) = \frac{ 2 k T}{\lambda^2} \int_{{\tiny LOS}}^{} \int_{a_{min}}^{a_{max}} \hspace{-0.6cm} \pi a^2 Q_{\rm abs}(a) n_{\rm dust}(a,r) n(a) da dz.
\end{equation}

Visibilities $\mathcal{V}$ as a function of $uv$-radius $r_{uv}$ are then
computed from this brightness distribution, taking into account
the finite size (with gaussian shape) of the primary beam
\citep[see][]{boissier2007,dbm2009}. The optical properties of the
individual grains are supposed not to vary with time, though sublimation-induced variations  might be expected.   The
absorption efficiencies $Q_{\rm abs}(a)$ are computed as explained in
Sect.~\ref{sec:thermal}.

\begin{figure*}
\caption{ Model results as a function of dust velocity parameter
$V_0$ for astronomical silicates with 0.5 porosity and 48\% ice
fraction. The outburst onset is $t_0$ = 23.8 UT and the maximum dust size is $a_{max}$ = 100
mm (top figures ({\bf a}) and ({\bf c})) and $a_{max}$ = 1000 mm (bottom figures ({\bf b}) and ({\bf d})).
Results for size indexes of $q$ = --3 and $q$ = --4 are plotted
respectively with solid, and dotted lines, and full
circles, and crosses. {\bf Left}: visibility curves
and interferometric-flux characteristics: $\mathcal{V}_{\rm 27
Oct}/\mathcal{V}_{\rm 28 Oct}$ at $r_{uv}$ = 20 m (red) and 70 m
(blue); interferometric flux ratio $F_{\rm 27Oct}/F_{\rm 28Oct}$
(green); index $q_v$ of visibility curve on 27 October (black). The
measured values and their error bars are shown by dotted
horizontal lines, and vertical lines, respectively. {\bf Right}:
reduced $\chi^2$ from data fit: 27 October data (red), 28 October
data (blue), 27 and 28 October data (black). } \label{fig:bestfit}
\begin{minipage}[t]{9cm}
\resizebox{\hsize}{!}{\includegraphics[angle=-90, bb = 110 17 530 674]{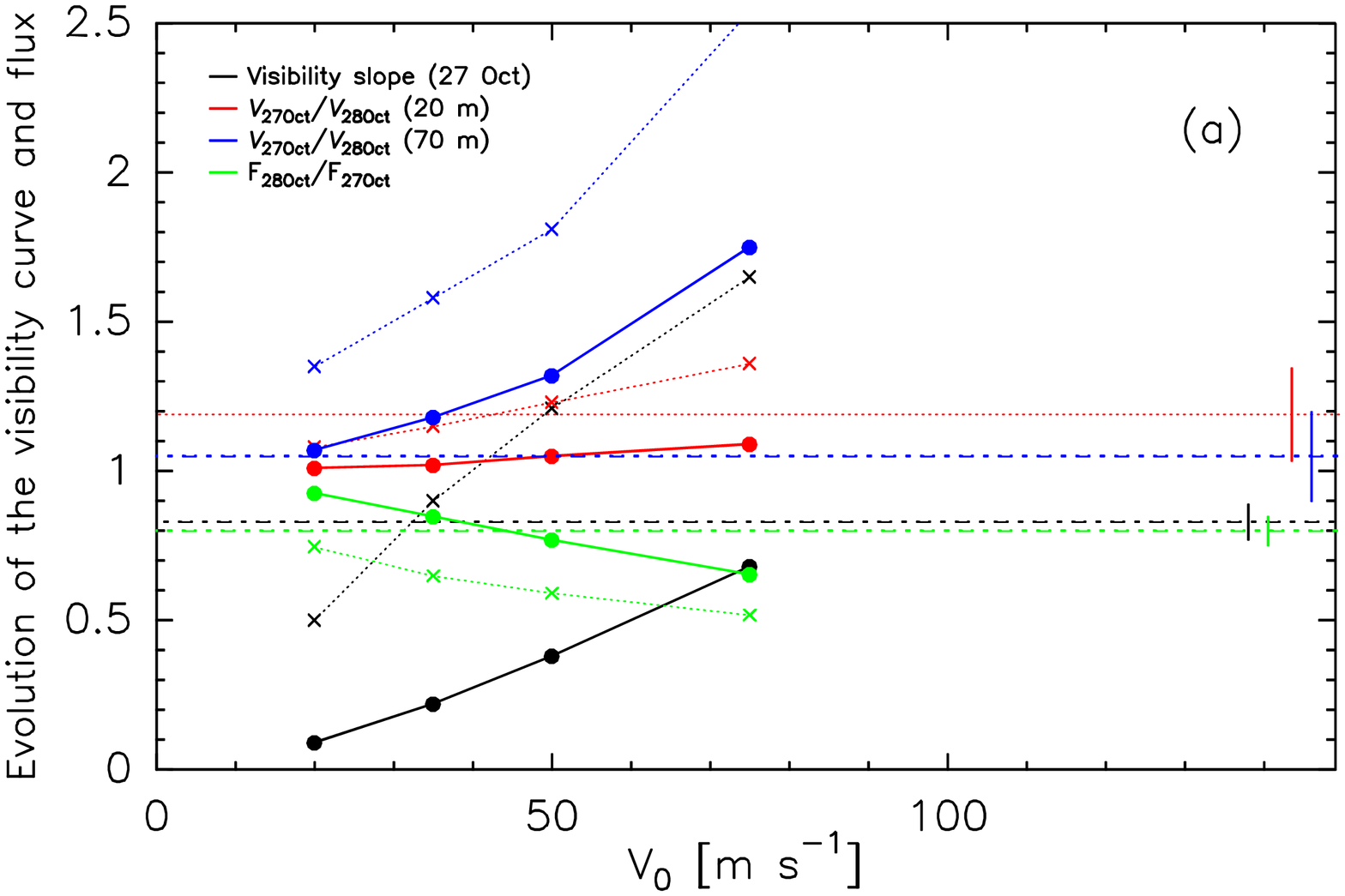}}
\end{minipage}\hfill
\begin{minipage}[t]{9cm}
\resizebox{\hsize}{!}{\includegraphics[angle=-90, bb = 110 17 530 674]{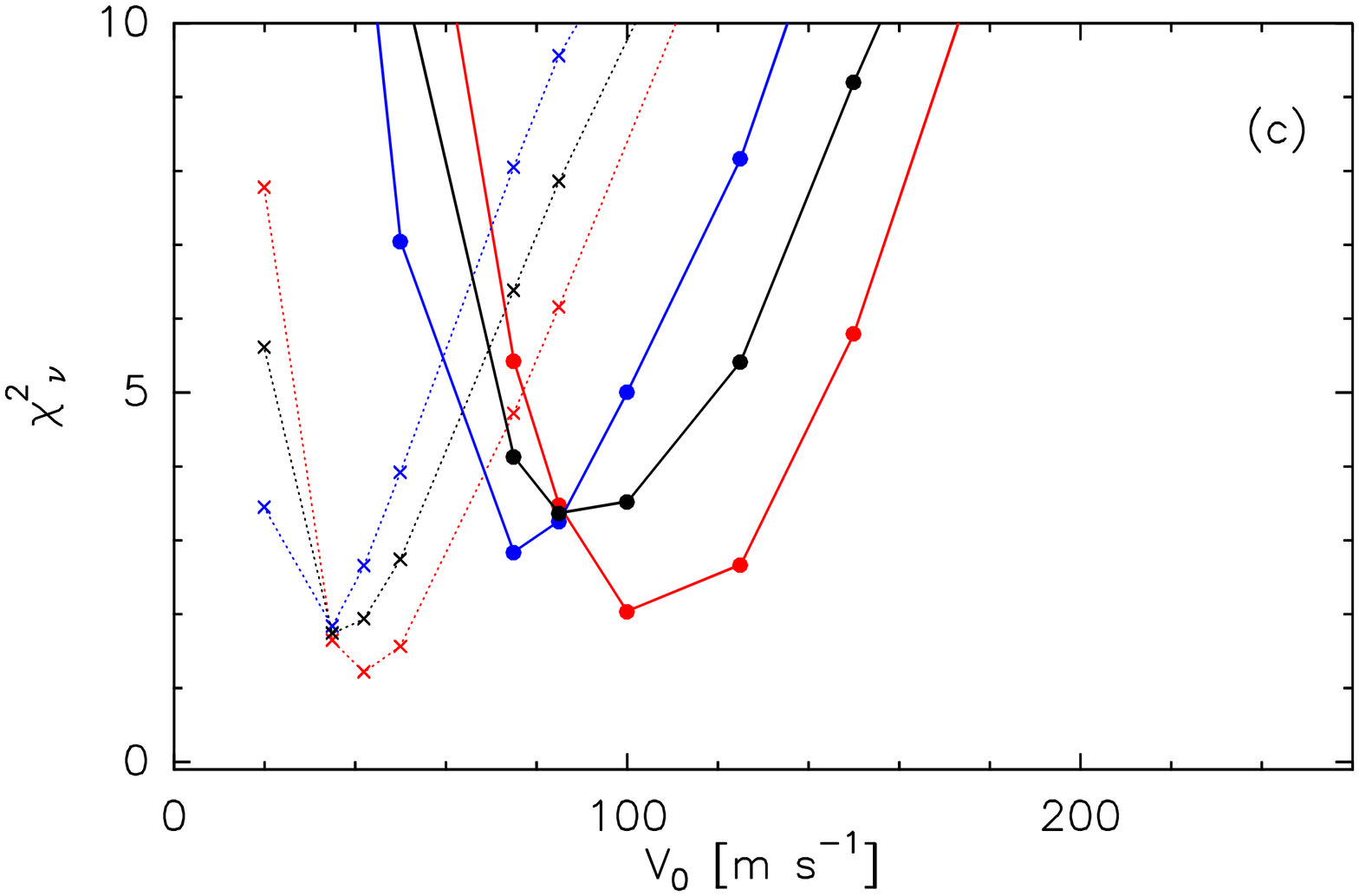}}
\end{minipage}
\begin{minipage}[t]{9cm}
\resizebox{\hsize}{!}{\includegraphics[angle=-90, bb = 110 17 530 674]{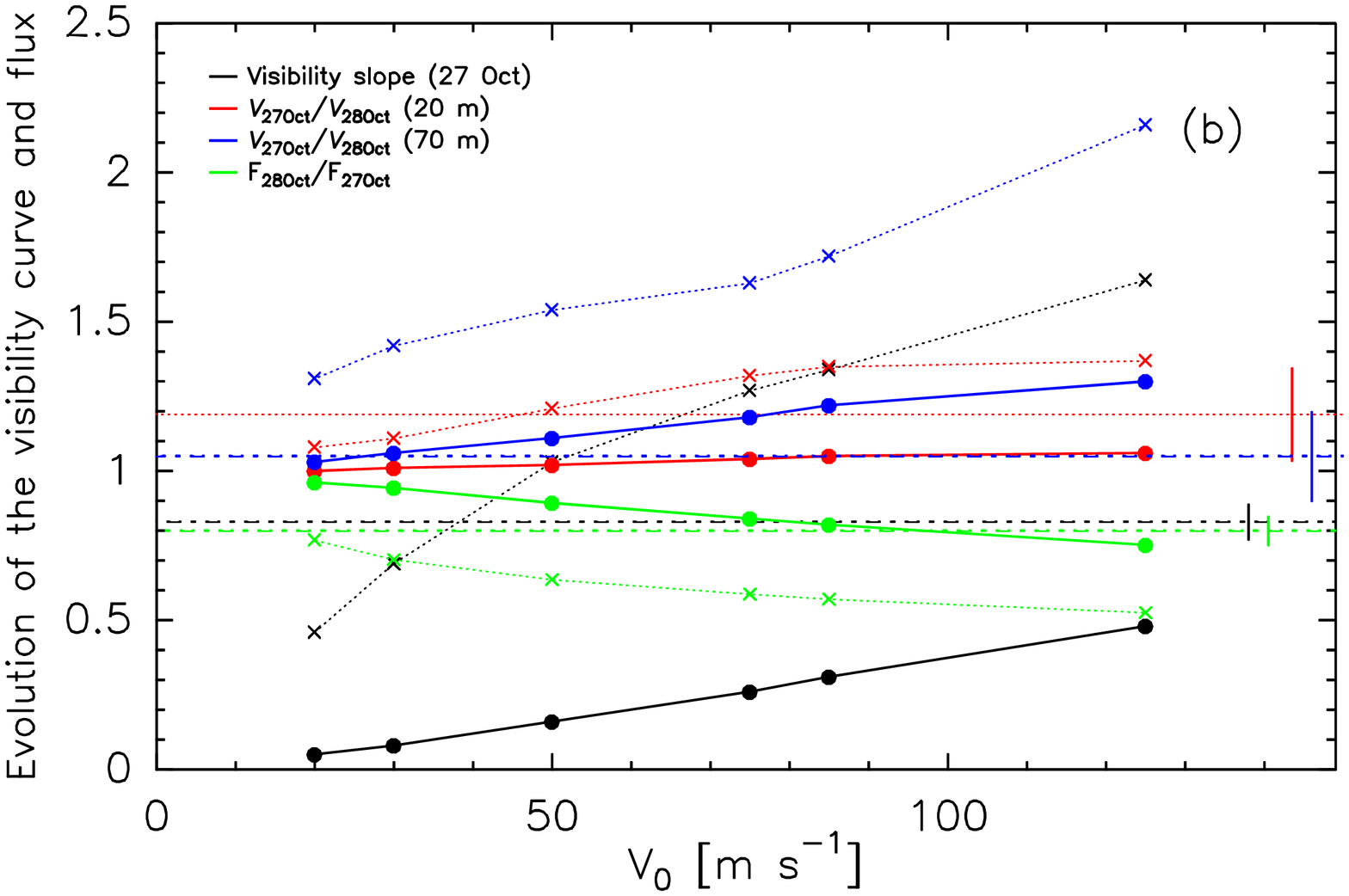}}
\end{minipage}\hfill
\begin{minipage}[t]{9cm}
\resizebox{\hsize}{!}{\includegraphics[angle=-90, bb = 110 17 530 674]{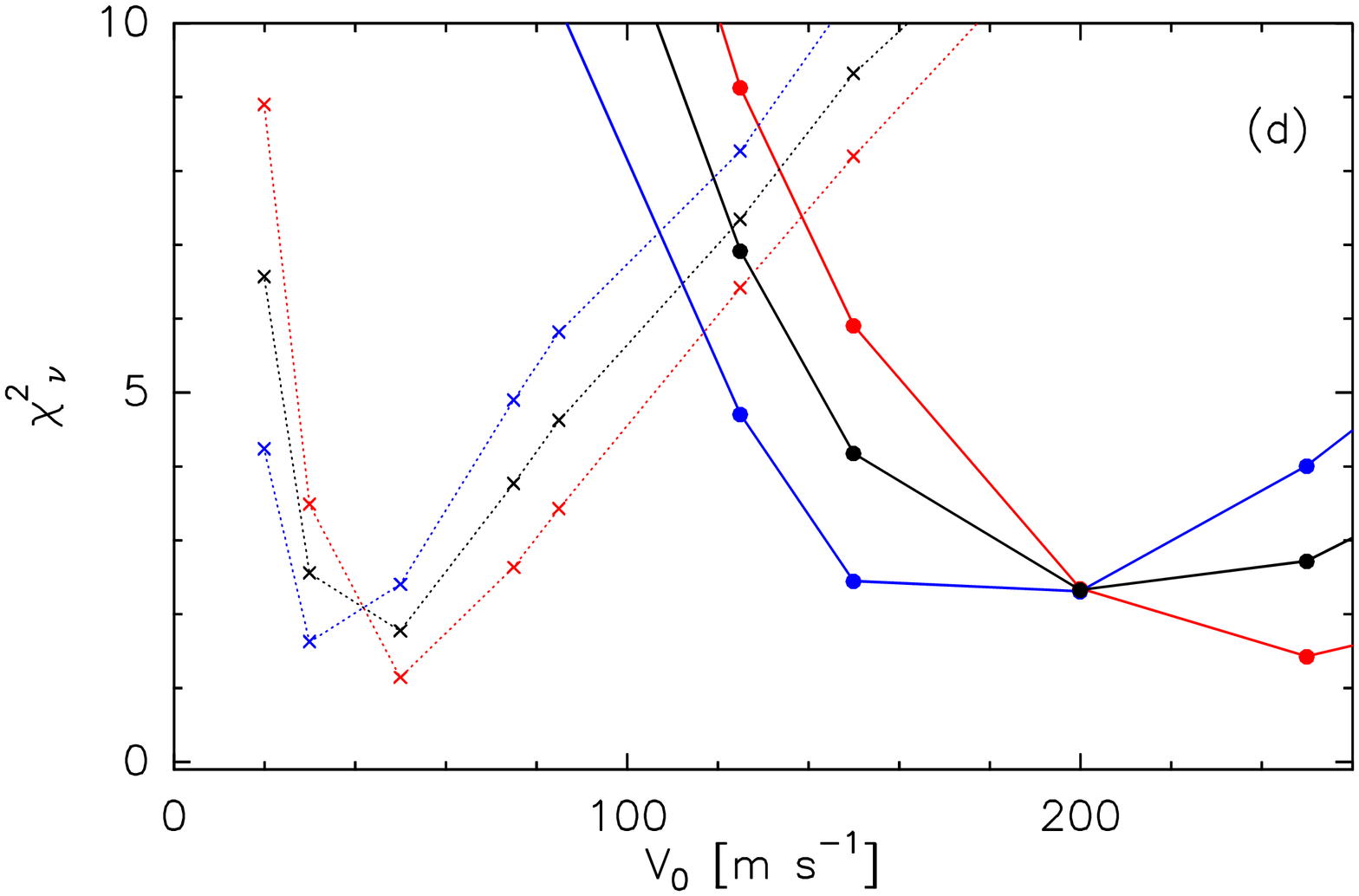}}
\end{minipage}
\end{figure*}

In our calculations, the number of bins in particle sizes is
typically 200--500. A high number of bins is a requisite for
getting a continuous dust spatial distribution when summing Eq.
\ref{eq:dust-density} over size. Actually, the number of bins is
fixed by the condition:
\begin{equation}
(V_{\rm dust}(i+1)-V_{\rm dust}(i))/V_{\rm dust}(i) <<
\Delta T_{\rm outburst}/(t-t_0)
\end{equation} 
\noindent
to be fulfilled for all bins i.

The velocity of the dust particles is a critical parameter. 
Information is available for the small particles, from the
evolution of the structure morphology of the expanding shell
observed in optical images. The expansion of the edge of the shell
is consistent with a velocity of $\sim$ 550 m s$^{-1}$
pertaining to particles in the 0.1--100 $\mu$m range \citep{Lin2009,Hsieh2010}. However, little
is known on the kinematics of large grains. Particles are likely
accelerated by their interaction with the flow of gas liberated by
sublimating icy grains. For steady-state comas, the terminal
velocity scales as $a^{-0.5}$ for intermediate sizes \citep{Crifo1997}. In order to roughly estimate
the range of expected velocities $V_0$ for 1-mm sized particles we
performed a 1-D dust dynamics modelling specific to comet
17P/Holmes using gas-drag coefficient for free molecular
flow. Nucleus gravity is computed assuming a nucleus density of 1000 kg m$^{-3}$. Gas dynamics is treated separately and assumes isentropic expansion. 
The gas total production rate was taken equal to 5 $\times$ 10$^{30}$ s$^{-1}$, based on the peak 
HCN production measured by \citet{biver2008} on  25.9 October UT and the HCN/H$_2$O relative abundance determined by  \citet{dello2008}. Under these assumptions, the terminal velocity of 1--mm radius particles is found to be  70 m s$^{-1}$, with a size dependence for $a$ = 100 $\mu$m to 10 cm close to $\delta$ = 0.5. The velocity of 1--100 $\mu$m particles ranges from 450 to 720 m s$^{-1}$, approaching the gas velocity (800 m s$^{-1}$) for the smallest sizes. The dynamics of 17P/Holmes's coma after the explosive event was obviously more complex than this simple steady-state model. Because gases outflow more rapidly than grains, the local gas pressure in the dust  ejecta cloud might have fastly declined with time, after the peak of gas production, reducing dust acceleration. For a gas production 10 times lower, the steady-state model predicts a terminal velocity for 1--mm particles $\sim$ 3 times lower. For the calculations, we have assumed $\delta$ = 0.5 and considered velocity values $V_0$ between 10 and 250 m s$^{-1}$.

\section{Results}
\label{sec:5}

\subsection{One-component models}
\label{sec:one-component}
Simulations were performed with an outburst of short duration
($\Delta T_{\rm outburst}$  = 0.3 d) compared to the elapsed time between outburst onset and 
PdBI observations (see further discussion at the end of Sect.\ref{sec:one-component}). We investigated several outburst onset times:  23.3 UT, 23.8 UT and 24.3 October UT. Indeed, though the onset time was established to be 23.3 UT by \citet{Hsieh2010},  this is somewhat disputed \citep[Z. Sekanina argues for an outburst occurring 0.4 d later, see note added in proof in][]{Li2011}.  
In addition, the maximum rate of brightening at optical wavelenths occurred on  24.5 October UT  \citep{Li2011}.  
 
The simulations were performed for astronomical silicates mixed with water ice (48\% in mass) and a porosity of 0.5. Figures~\ref{fig:bestfit}a--b 
plot, as a function of the velocity parameter $V_0$ and for
$\delta$ = 0.5 (Eq.~\ref{equ:4}), several quantities
characterizing the evolution of visibilities from  27 to
28 October: the ratios $\mathcal{V}_{\rm 27 Oct}/\mathcal{V}_{\rm 28 Oct}$
at $r_{uv}$ = 20 m and 70 m, and the ratio of the flux density
 within the interferometric beam $F_{\rm 28Oct}/F_{\rm 27Oct}$. Also plotted is the slope
index $q_v$ for 27 October, defined according to $\mathcal{V}_{\rm 27 Oct}$
$\propto r_{uv}^{-q_v}$. The observed values of these quantities are also given in the figure. 
We see that there is a correlation
between $q_v$ and the decrease of $F$ and $\mathcal{V}$ from
 27 to 28 October. Larger grain velocities lead to a faster
decrease of the flux density and to a more extended coma, which,
in the Fourier plane, results in a faster decline of $\mathcal{V}$
with $r_{uv}$ (i.e., large $q_v$, Fig.~\ref{fig:bestfit}).
Alternatively, for low velocities, the dust coma remains unresolved for most baselines 
and $q_v$ approaches the zero value characterizing an unresolved source. For a large enough dust
parameter $V_0$, the material contributing to the 3.3 mm emission forms
a ring of particles which inner boundary has a size comparable or larger than the interferometric beam, and
the real part of the visibility becomes negative at large $r_{uv}$. This 
is unlike the measurements, implying that low-velocity particles 
present near the nucleus are detected with  the longest baselines at the two dates.  
Model results shown in Fig.~\ref{fig:bestfit} consider maximum size 
particles $a_{max}$ equal to 100 and 1000 mm. When $a_{max}$ is set
to 10 mm, only a small size range is contributing to the 3.3 mm radiation (Fig.~\ref{fig:para}). 
This thin shell of radiating grains rapidly moves through the 
field of view. Visibility curves then display either a small $r_{uv}$ dependence (unresolved shell, low $q_v$) or the shape characteristics of a ring-like structure explained above (steep  $r_{uv}$ dependence and real part with negative values at large $r_{uv}$). For $t_0$ = 23.8 h,
the transition between the two cases is obtained for $V_0$ $\sim$ 20--30 m s$^{-1}$ (consistent with $V_0 \times (t-t_0) \sim$ FOV radius). 
     
\begin{figure}[h]
\resizebox{9cm}{!}{\includegraphics[angle=-90, bb = 110 17 530 674]{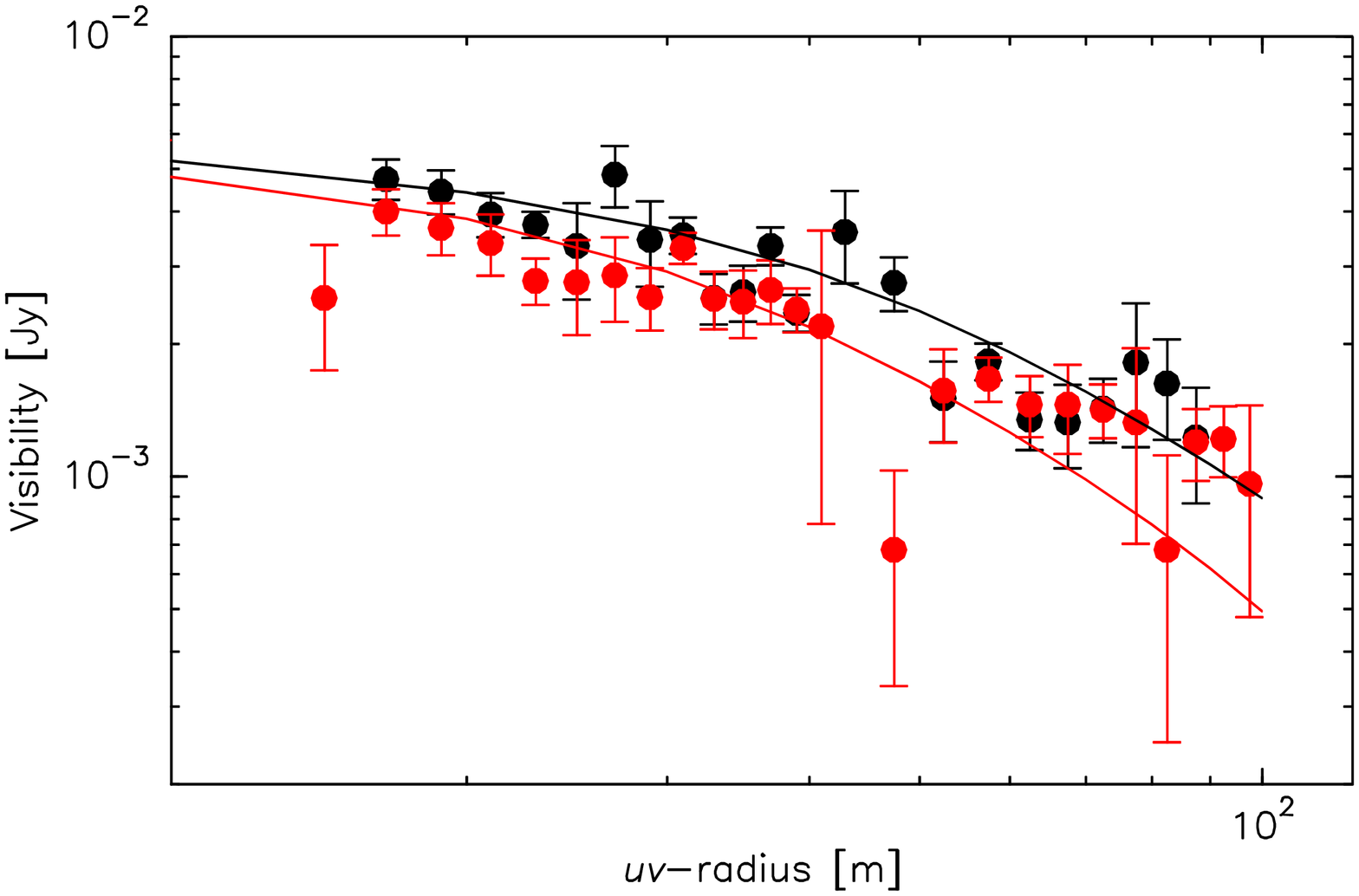}}
\caption{Visibility curves with the one-component model, for an outburst with $\Delta T_{\rm outburst}$ =
25000 s, a size distribution in $a^{-4}$ with $a_{max}$ = 100 mm, 
 a velocity law $V_{\rm dust}(a) = V_0~(a/a_0)^{-0.5}$ with $V_0$ = 35 m s$^{-1}$ being the velocity of 1-mm particles, and an outburst onset $t_0$ = 23.8 UT. Data are shown by filled circles with error bars. The solid curves show the model results. Black
(red) symbols and curves are for  27 and 28 October, respectively.
Dust particles are made of a porous (50\% porosity) mixture of
astronomical silicates and ice, with an ice mass fraction of 48\%.}
\label{fig:out1}
\end{figure}

The observed visibilities were compared to the modelled curves. 
The resulting reduced $\chi^2_\nu$ for the 27 and 28 October data, and
for the fit of both 27 and 28 October data, are shown in
Fig.~\ref{fig:bestfit}c--d for $t_0$ = 23.8 h and $a_{max}$ = 100 and 1000 mm. The velocities $V_0$ providing the best
$\chi^2_\nu$ are given in Table~\ref{result}. $V_0$ is lower for a
steeper size distribution because the 3.3 mm emission is then
sampling a relatively larger amount of small dust particules with larger velocities. For the same reason, $V_0$ decreases
with decreasing $a_{max}$ (Table~\ref{result}). On the other hand, retrieved $V_0$
are not very sensitive to the outburst onset time for $t_0$ within 23.3--24.3 UT. For 
example, for $a_{max}$ = 100 mm, $q$ = --3/--3.5/--4, , $V_0$ is $\sim$ 75/60/40, 85/65/35, and 100/85/40 m s$^{-1}$ for $t_0$ = 23.3, 23.8, and 24.3 October UT respectively.  
Altogether, $V_0$ values range from
typically $\sim$ 40 to 100 m s$^{-1}$, with values as large as 200 
m s$^{-1}$ being obtained for the size distribution in $a^{-3}$ (Table~\ref{result}).
Such high values are likely unrealistic as the velocities for 1-mm grains would then be comparable to the velocities measured for
$\mu$m-sized grains from optical data \citep{Lin2009,Hsieh2010}.
As indicated by the reduced $\chi^2_\nu$, the observations are
best fitted for steep size distributions (Table~\ref{result}),
with $V_0$ $\sim$ 30--40 m s$^{-1}$. Figure~\ref{fig:out1} shows 
an example of model fits for a size distribution in $a^ {-4}$.

While satisfactory fits are obtained for
the visibility curve obtained on 27 October ($\chi^2_\nu$ $\sim$ 1.2--1.5),
the fit for 28 October is poorer (Figs.~\ref{fig:bestfit}, \ref{fig:out1}).
In fact, this one-component model fails in explaining the small temporal evolution of
$\mathcal{V}$(70 m) (see Fig.~\ref{fig:out1}) and the measured ratio $F_{\rm 28Oct}/F_{\rm 27Oct}$ of 0.80 $\pm$ 0.05 (see Table~\ref{result}). Since replenishment
of the coma (e.g., by fragmentation processes) proceeded after outburst onset, we could question the production function $G(t)$ used in our model. \citet{Li2011} observed that the 
rate of change of the scattering cross-section followed a Gaussian-like function of $\sim$ 0.4 d width and peaking on 24.54 October UT. Our $G(t)$ function has the same width  (note that model results are not much sensitive to the duration of injection of material, provided this duration remains small with respect to the elapsed time between outburst onset and the measurements). Model calculations with 
$t_0$ $\simeq$ 24.5 October UT do not explain the measurements. In other words, the production of the 3-mm radiating particles did not proceed as for the micron-sized particles. The longest baselines probe typically distances from the nucleus $r \leq $ 4000 km. The constancy of $\mathcal{V}$ for $r_{uv}$ $>$ 60 m over Oct. 27--28 suggests that the outburst produced a second population of very slowly
moving grains which 3-mm emission remained unresolved by the interferometric beam on both dates. This population of slowly moving grains produced by the outburst has been identified in mid-infrared images and referred as to the "core" component \citep{reach10}. Its origin will be discussed in Sect.~\ref{sec:6}. In the next section, our model will incorporate this second population of debris.

\subsection{Two-component models}

\begin{table*}[t]
\begin{center}
\caption{Two-component model: best fit for astronomical silicates with 0.5
porosity and 48\% ice fraction.} \label{result2}
\begin{tabular}{l | l l c c c c | c | c}
\hline \noalign{\smallskip} \hline \noalign{\smallskip}
&&&&Shell&&&Core & \\
\hline\noalign{\smallskip}
Onset & Size  & $a_{max}$ & $V_0$ & $M_{\rm dust}^{\rm beam (a, b)}$ & $M_{\rm dust}^{\rm tot (b, c)}$  & $C_{\rm dust}^{\rm tot (d)}$ & 3-mm Flux & $\chi^2_\nu$ $^{(e)}$\\
\noalign{\smallskip}
 (October UT) & index & (mm) & (m s$^{-1}$) & (10$^{11}$ kg) & (10$^{11}$ kg) & (10$^{12}$ m$^2$) & (mJy) & \\
\hline\noalign{\smallskip}
23.3 & $-$3.5 & 100 & 125 & 1.6 & 5.5 & 1.8 -- 5.8 & 1.41 $\pm$ 0.05 & 1.3 \\ 
 & $-$4.0 & 100 & 75      & 1.4 & 9.8 & 89 -- 873 & 1.26 $\pm$ 0.05 & 1.4 \\ 
23.8 & $-$3.5 & 10 & 60   & 0.60 & 3.2  & 3.3 -- 10 & 1.60 $\pm$ 0.05  & 1.6\\
 & $-$4.0 & 10 & 50       & 0.70 & 8.4 & 95 -- 924 & 1.57 $\pm$ 0.05 & 1.6 \\   
 & $-$3.5 & 100 & 125     & 1.7 & 5.0 & 1.6 -- 5.2 & 1.27 $\pm$ 0.05 & 1.3 \\ 
 & $-$4.0 & 100 & 85      & 1.4 & 9.7 & 87 -- 861 & 1.27 $\pm$ 0.05 & 1.4 \\ 
 & $-$3.5 & 1000 & 250    & 8.3 & 16 & 1.7 -- 5.4 & 1.13 $\pm$ 0.05 & 1.3\\ 
 & $-$4.0 & 1000 & 75     & 2.9 & 9.7 & 70 -- 715 & 0.87 $\pm$ 0.05 & 1.3\\
\hline
\end{tabular}
\end{center}
$^{(a)}$ Dust mass within the synthesized beam on 27 October. \\
$^{(b)}$ For $a_{min}$ = 1 $\mu$m. Values obtained with $a_{min}$ = 0.1 $\mu$m differ by less than 20\%.\\
$^{(c)}$ Total dust mass in the shell produced by the outburst.\\
$^{(d)}$ Total scattering cross-section for $a_{min}$ in the range 0.1--1 $\mu$m.\\ 
$^{(e)}$ Reduced $\chi^2$ from the fit of 27 and 28 October data.\\

\end{table*}

\begin{figure*}[t]
\begin{minipage}[t]{9cm}
\resizebox{9cm}{!}{\includegraphics[angle=-90, bb = 110 17 530 674]{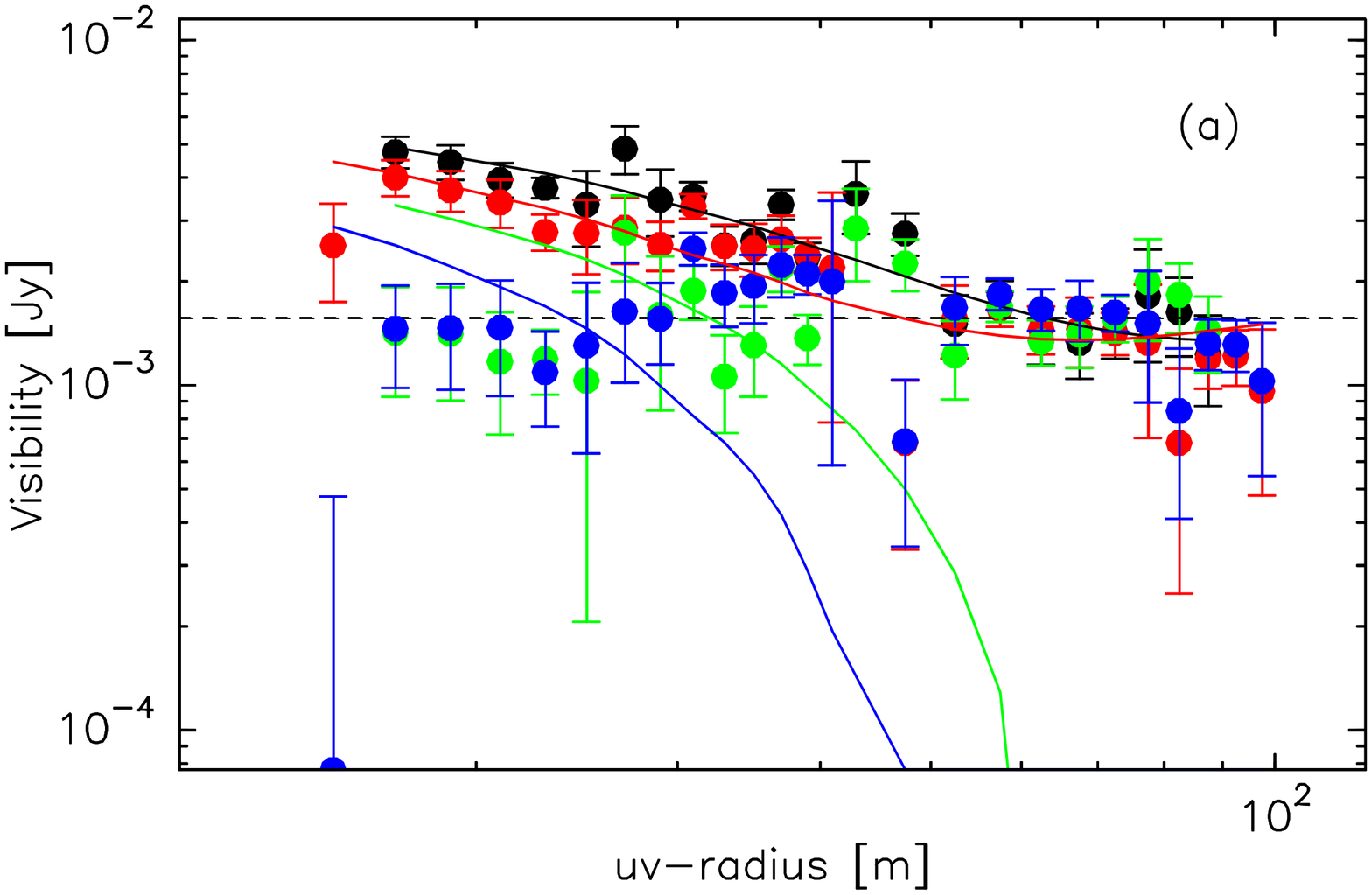}}
\end{minipage}\hfill
\begin{minipage}[t]{9cm}
\resizebox{9cm}{!}{\includegraphics[angle=-90, bb = 110 17 530 674]{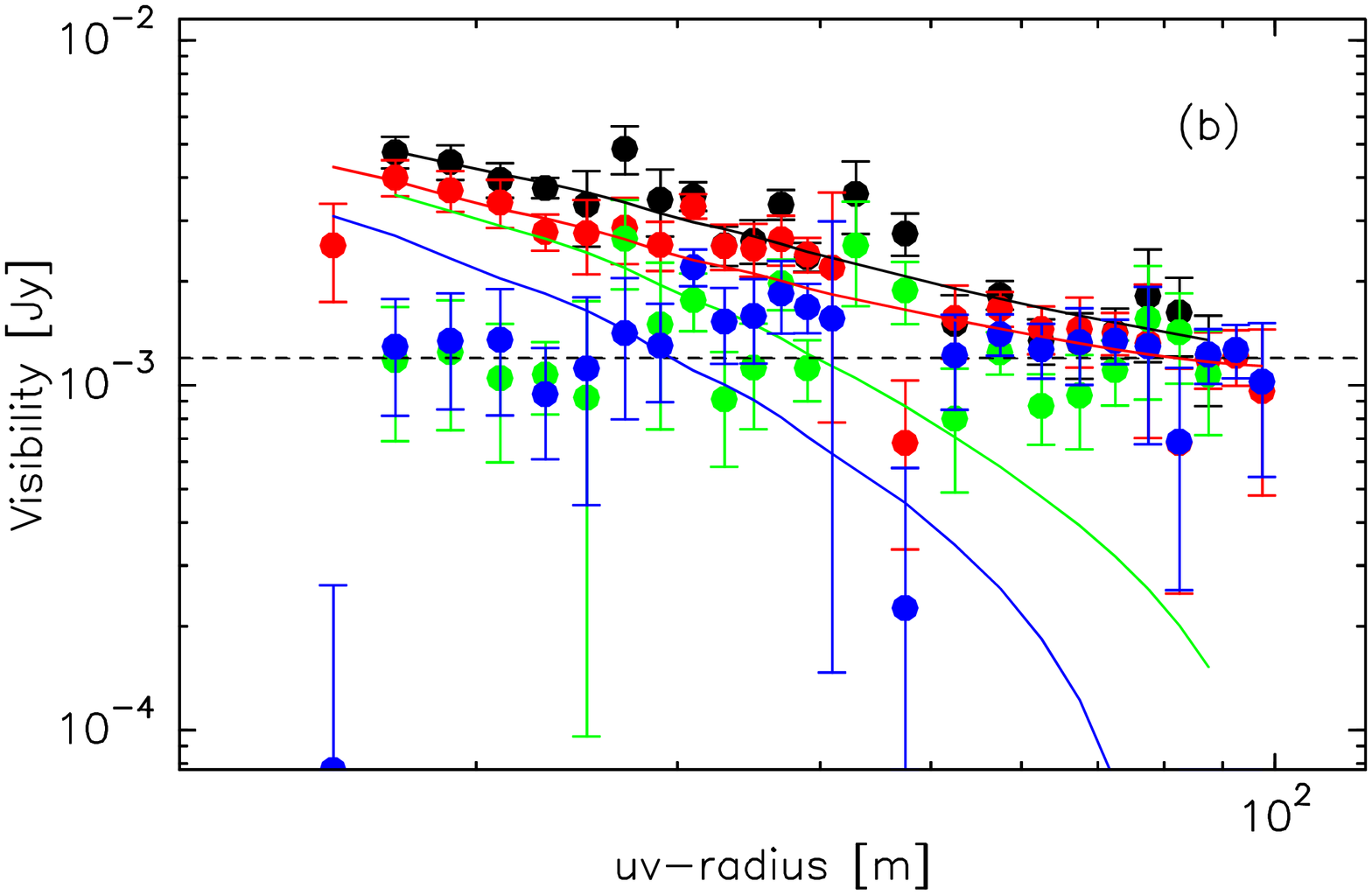}}
\end{minipage}

\caption{Visibility curves for the two-component model. Model curves from the shell emission and residuals (data  minus shell emission) are shown by a green (blue) line and green (blue) plain circles, respectively, for  27 (28) October UT. Parameters for the shell emission are: outburst with $\Delta T_{\rm outburst}$ = 25000 s, $t_0$ = 23.8 UT, and a size distribution in $a^{-4}$: {\bf (a)} $V_0$ = 50 m s$^{-1}$, $a_{max}$ = 10 mm; {\bf (b)} $V_0$ = 85 m s$^{-1}$, $a_{max}$ = 100 mm. The point source (core) emission is shown with the dashed black line. The sum of the shell and core emissions are shown in solid lines. Data are shown by filled circles with error bars. Black
(red) symbols and curves are for  27 and 28 October, respectively.
Dust particles are made of a porous (50\% porosity) mixture of
astronomical silicates and ice, with an ice mass fraction of 48\%.}
\label{fig:outdiff}
\end{figure*}

\begin{figure}[b]
\resizebox{9cm}{!}{\includegraphics[angle=0]{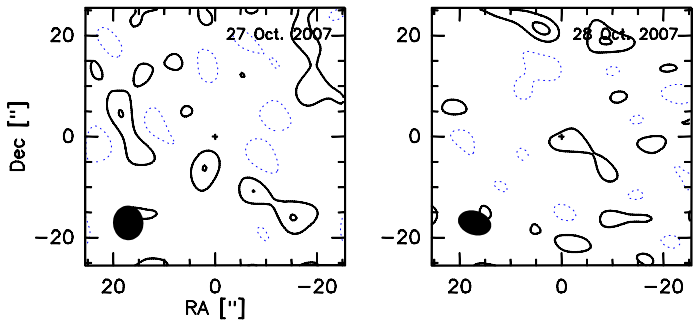}}
\caption{Residuals obtained from the difference between observations and the two-component model. Model parameters are those of Fig.~\ref{fig:outdiff}b. Left: residuals for 27 October. Right: residuals for 28 October.  The levels are 1$\sigma$ spacing with $\sigma$ = 0.093 and 0.12 mJy/beam for October 27 and 28, respectively. The synthesized interferometric beam is plotted in the left corner.}
\label{fig:resid}
\end{figure}

In order to characterize the unresolved population of slowly moving grains, we fitted the observations by the 
combination of emission of outflowing dust grains (hereafter refered as "shell"), and a point-source emission (the "core" component).  The shell emission is modeled as described in Sects.~\ref{sec:model} and \ref{sec:one-component}. The flux density from the core was determined from the constraint that the core remained unresolved on  27 and 28 October. Figures~\ref{fig:outdiff}a--b show, for two sets of parameters, the modelled visibilities for the two components and the comparison to the measurements. At $r_{uv}$ $>$ 60 m, the ratio $\mathcal{V}_{\rm 27Oct}$/$\mathcal{V}_{\rm 28Oct}$ is 1--1.2, consistent with the measured value of 1.05 $\pm$ 0.15 (see Fig.~\ref{fig:bestfit}). The best fits are obtained for grain velocities in the shell higher than determined from the one component model. The results of this two-component modelling are summarized in Table ~\ref{result2}.   
 The flux density from the core is in the range 0.9--1.6 mJy. The highest values are obtained when $a_{max}$ = 10 mm. In this case, the 3.3 mm-radiating shell has a small thickness and its inner boundary is larger than the interferometric beam, so that most of the emission    
detected with the long baselines is from the material composing the core (Fig.~\ref{fig:outdiff}a).  Residuals between modelled and observed maps are shown in Fig.~\ref{fig:resid} for the set of model parameters of Fig.~\ref{fig:outdiff}b.

\begin{figure}[t]
\resizebox{10cm}{!}{\includegraphics[angle=-90]{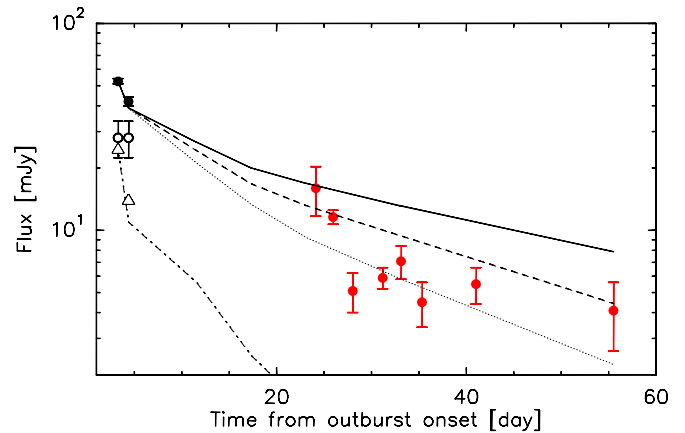}}
\caption{Evolution of the 250 GHz flux from IRAM 30-m telescope \citep[red dots,][]{alten2009} and PdBI data (black dots). PdBI 90-GHz fluxes were converted to 250 GHz fluxes, applying a conversion factor of 22.1 determined by thermal modelling. The beam size is $HPBW$ = 11$\arcsec$ for the 30-m data and $\sim$ 6$\arcsec$ for the PdBI data. The observed core 
 (Table~\ref{result2}) and shell emissions on  27--28 October are shown by opened circles and triangles, respectively. The dotted-dashed line shows the 
model for the shell emission ($q$ = --4, $a_{max}$ = 100 mm, $V_0$ = 85 m s$^{-1}$) with $HPBW$ = 6$\arcsec$ for the two first dates, and  11$\arcsec$ later on. Two-component models (shell+core) are shown with solid, dashed and dotted lines, with the core parameters being $q$ = --3.2 and ($a_{max}$ = 100 mm, $V_0$ = 5 m s$^{-1}$), ($a_{max}$ = 100 mm, $V_0$ = 9 m s$^{-1}$), and ($a_{max}$ = 10 mm, $V_0$ = 7 m s$^{-1}$), respectively.     }
\label{fig:30m}
\end{figure}

The PdBI observations do not provide strong constraints on the kinematic properties of the dust particles composing the core. As already discussed in Sect.~\ref{sec:2}, the mean velocity $V_{\rm mean}^{\rm 3mm}$ of the core should not exceed 10 m s$^{-1}$ to remain unresolved by the PdBI on the two dates. Instead, we used published continuum measurements obtained at 250 GHz (i.e., 1.1 mm wavelength) using the IRAM 30-m telescope ($HPBW$ = 11$\arcsec$) which cover the 16 November to 18 December period \citep{alten2009}. Absorption coefficients were computed at 1.1 mm, following Sect.~\ref{sec:thermal}. Figure~\ref{fig:30m} shows the measurements of the dust continuum flux at 1.1 mm, where the 3.3 mm flux densities recorded at the PdBI were extrapolated to 1.1 mm, based on our calculations of dust opacities at the two wavelengths. Because they were performed late after the outburst, the IRAM 30-m telescope detected mainly emission from the low-velocity particles composing the core (see the modelled time evolution of the shell emission in Fig.~\ref{fig:30m}). This is in agreement with the conclusion obtained by \citet{reach10} combining IRAM and {\it Spitzer} data sets. The time evolution of the core emission depends on the particle size distribution and velocity field. Examples of model fits for $q=$ --3.2 (Sect.~\ref{dust-mass}) are given in Fig.~\ref{fig:30m}. The velocity of 1-mm particles in the core is typically $V_{0}$ = 7--9 m s$^{-1}$. The velocity of the largest particles (0.1 m) is on the order of the escape velocity (estimated to 0.8 m s$^{-1}$).  From mid-infrared images of the core obtained in March 2008, \citet{reach10} inferred a velocity of 9 m s$^{-1}$. The agreement with our result is excellent.

\subsection{Dust masses and size indexes}
\label{dust-mass}
The time-dependent model allows us to compute not only the mass of dust particles
within the field of view at the time of the measurements $M_{\rm dust}^{\rm beam}$, but also the total mass injected by the outburst $M_{\rm dust}^{\rm tot}$. These quantities are given in Tables~\ref{result} and~\ref{result2} for the one-component and two-component models, respectively. For the shell, $M_{\rm dust}^{\rm tot}$ differs from $M_{\rm dust}^{\rm beam}$ measured on 27 October by a factor of 2 to 12, the largest difference being obtained for the size distributions with size index $q$ = --4 (Table~\ref{result2}). Most of the initial mass is then comprised in small, rapidly moving particles which are outside the PdBI field of view on  27 October. The dust mass found in the shell component ranges from 3 to 16$\times$ 10$^{11}$ kg (Table~\ref{result2}). However, models requiring velocities as large as 250 m s$^{-1}$ for 1-mm particles can be excluded (Sect.~\ref{sec:model}). Therefore, values in the range 3--10 $\times$ 10$^{11}$ kg are more likely, under the assumption that the optical properties assumed for our calculations are representative of comet grains.

For comparison with optical data, Tables~\ref{result} and~\ref{result2} present total dust scattering cross-sections $C_{\rm dust}^{\rm tot}$ inferred for the different models, assuming minimum particle sizes in the range $a_{min}$ = 0.1--1 $\mu$m. Particles with sizes less than 0.1 $\mu$m are not considered as they do not scatter efficiently visible light. The scattering cross-section measured from optical observations reached  5.5 $\times$ 10$^{13}$ m$^2$ at the peak of brightening  on $\sim$ 25.0 October UT \citep{Li2011}.  For $a_{min}$ = 1 $\mu$m, the  best fit for the two-component  model implies $C_{\rm dust}^{\rm tot}$ = 1.6--3.3 $\times$ 10$^{12}$ m$^2$ for $q$ = --3.5, and 
$C_{\rm dust}^{\rm tot}$ = 70--95 $\times$ 10$^{12}$ m$^2$ for $q$ = --4 (Table~\ref{result2}), which suggests a size distribution in the shell component close to $a^{-4}$ if the minimum size is 1 $\mu$m. From interpolation, we found that the size index fitting both optical and radio data is $q$ $\sim$ --3.7 for $a_{min}$ = 0.1 $\mu$m and --3.9 for $a_{min}$ = 1 $\mu$m. 
These values are obtained for astronomical silicates mixed with water ice and a porosity of 50\%.
Taking into account the range of grain compositions and porosities considered in Table~\ref{tab:opacity}, $q$ is within [--3.8, --3.6] and [--4.1, --3.8], for $a_{min}$ = 0.1 and 1 $\mu$m, respectively. The inferred size index is not significantly sensitive to $a_{max}$.
 However, our model is not considering time-dependent dust production processes, such as particle fragmentation, responsible for the rapid increase of the scattering cross-section after outburst onset \citep{Hsieh2010,Li2011}. This introduces additional uncertainties in the size index determination.

The size distribution in the core component can be evaluated from optical data contemporaneous to PdBI data.
We assumed that the bright condensation surrounding the nucleus seen in optical images is the visible counterpart of this core component. From images obtained by N. Biver and J. Nicolas (private communication) on  26.9 and 27.8 October UT, the scattering cross-section in a 6$\arcsec$ diameter aperture corresponding to the PdBI field of view was 0.7 and 0.4\% the total scattering cross-section, implying a scattering cross-section of at most $\sim$ 3.5 $\times$ 10$^{11}$ m$^2$ in the core on  27.1 October UT. The small variation of the cross-section from  27 to 28 October can be interpreted as indicative of slowly moving particles ($\sim$ 15 m s$^{-1}$, if we assume that dust production from the nucleus or from fragmentation was negligible between the two dates).  The $Q_{\rm abs}$-weighted cross-section of 3 mm-radiating particles derived from the 1.3 mJy flux density (Table~\ref{result2}) is 1.8 $\times$ 10$^{9}$ m$^2$. The comparison with the scattering cross-section shows that the core contained a larger proportion of large particles, compared to the shell.  Assuming $a_{min}$ = 0.1 $\mu$m and          
$a_{max}$ = 10 to 1000 mm, optical and radio data can be reconciled with a size index of --3.2 to --3.0. The results are not much sensitive to the minimum particle size: $q$ is within [--3.3, --3.0] for $a_{min}$ = 1 $\mu$m. Using these size indexes, we derive dust masses for the core of 0.7, 1.2, and 3.7 ($\times$ 10$^{11}$) kg for $a_{max}$ = 10, 100, and 1000 mm, respectively. Optical and radio data cannot be both explained for maximum grain sizes set to $<$ 2 mm. These calculations pertain to the nominal composition (50\% astronomical silicate, 48\% ice, porosity of 0.5). Considering the range of dust composition and porosity in Table~\ref{tab:opacity}, $q$ remains within [--3.35, --3.0] for $a_{min}$ = 0.1 $\mu$m. Given uncertainties in the size distributions for the two components, especially on $a_{max}$, the ratio $M_{\rm dust}^{\rm core}/M_{\rm dust}^{\rm shell}$ is in the range 0.1--1.

In summary, we derived dust masses of 3--10 $\times$ 10$^{11}$ kg and 0.7--4 $\times$ 10$^{11}$ kg for the shell and core components, respectively. These calculations are valid for astronomical silicates mixed with water ice, with a porosity of 0.5. Additional uncertainties come from the optical properties of the dust ejecta which are not precisely known. Based on our calculations of dust opacities for different materials (Table~\ref{tab:opacity}), dust masses could be 60\% higher (material referred as to organics and silicate mixture in the table) or up to a factor 2.6 lower (higher porosity, ice-free).

\section{Discussion}
\label{sec:6}

We performed a detailed analysis of the 3.3 mm dust thermal emission from comet 17P/Holmes observed 4--5 days after its massive outburst with the IRAM Plateau de Bure interferometer. The data were combined to optical and IRAM 30-m telescope measurements to characterize the properties of the ejecta cloud, namely the dust size distribution, its kinematics and mass.

Two distinct dust components, with different kinematic properties and size distributions, are identified in the data. The large-velocity component, with typical velocities $V_0$ of 50--100 m s$^{-1}$ for 1 mm particles,  corresponds to the fast (550 m s$^{-1}$) expanding shell observed in optical images and displays a steep size distribution with a size index estimated to $q$ = --3.7. The size dependence of the velocity is consistent with gas drag (Sect.~\ref{sec:model}). The very high gas production rates measured shortly after the outburst ($>$ 10$^{30}$ s$^{-1}$) and their fast temporal drop--off argue for gas production from subliming icy grains \citep{biver2008,combi2007,sch2009}. High gas pressures caused the particles constituting the shell to be accelerated to large terminal velocities. The second component, the "core", consists in slowly-moving particles with kinematic properties ($V_0$ = 7--9 m s$^{-1}$) consistent with results obtained by \citet{reach10} from the study of mid-infrared images. Constraints obtained on the size index and maximum dust size show that the core is dominated by large ($a_{max}$ $>$ 2 mm) particles with a shallow size index $|q|$ $<$ 3.4. The same conclusion was obtained by \citet{reach10} from the analysis of the {\it Spitzer} images. In addition, the core revealed featureless 10-$\mu$m spectra in mid-November 2007, in contrast to the shell, consistent with much larger grains populating the core \citep{reach10}. 

The origin of the low-velocity core component may be related to the fast decline of the gas pressure with time in the inner coma resulting from the fast \citep[$>$ 1 km s$^{-1}$,][]{boissier2009} expansion of the first generation of gases. Under such conditions, we expect gas velocities to vary both with time and distance to nucleus, with the older molecules present at larger distances reaching higher velocities, and this was indeed observed \citep{biver2008,boissier2009}. The fast decline of gas pressure and velocity in the inner coma resulted in a strong decrease of the gas aerodynamic force acting on grains, making their acceleration less effective. Grains produced after the peak of gas production were thus accelerated to lower terminal velocities. Another effect to consider is that efficient grain acceleration requires gas pressure gradient. Particles in the centre of the expanding cloud were subject to limited acceleration since the gas pressure gradient caused by the subliming grains was minimum in this region. Finally, slowly-moving particles, in the trailing side of the ejecta cloud, were decelerated due to the positive gradient of pressure in the radial direction combined with nucleus gravity. The kinematics of the most massive debris was possibly not significantly affected by gas-drag; providing they were not accelerated by rocket forces, their velocity remained close to their separation velocity from the nucleus.  Thus, the low velocity of the particles and debris composing the core might have different origins.

We derived dust masses of 3--10 $\times$ 10$^{11}$ kg and 0.7--4 $\times$ 10$^{11}$ kg for the shell and core, respectively, using grains made of astronomical silicates mixed with water ice as cometary analogues. The ratio $M_{\rm dust}^{\rm core}/M_{\rm dust}^{\rm shell}$ is in the range 0.1--1, lower than the value $\sim 2$ determined by \citet{reach10}. The large mass in the shell component and its steep size distribution indicate that the disruption of the nucleus was accompanied or followed by a massive production of small particles, indicative of material with small cohesive strength.  

We found that the total dust mass $M_{\rm dust}$ injected by the outburst was in the range 4--14 $\times$ 10$^{11}$ kg. This is in the high end of reported values based on optical or infrared observations. From optical observations, \citet{Li2011} estimated $M_{\rm dust}$ to 0.2--9 $\times$ 10$^{11}$ kg , while \citet{sch2009} and \citet{sekanina2008} favour values of 1--2 $\times$ 10$^{11}$ kg. The estimate made by \citet{reach10} from their {\it Spitzer} thermal observations is 0.1 $\times$ 10$^{11}$ kg. The inconsistency between these concurrent measurements reflects different assumptions on the size distribution, especially on the maximum particle size 
in the ejecta. For example, \citet{reach10} made their estimations using characteristic sizes in the core, blob and shell components of 200, 8 and 2 $\mu$m, respectively, so that their values are lower limits to the dust masses. 

The total dust mass is about 10 times higher than the integrated water production of $\sim$ 5 $\times$ 10$^{10}$ kg over several months \citep{biver2008,sch2009}. Most of the water was produced on a timescale of the order of the day after outburst onset from small dirty icy grains. On-going gas production was observed through March 2008  \citep{sch2009}. The spatial distribution of OH radicals suggests outgassing from the inner part of the coma, typically $<$ 1--2 $\times$ 10$^4$ km according to \citet{sch2009}. A significant contribution from 
the icy debris in the core is likely, though detailed modelling is required to investigate whether this interpretation is consistent with the observed radial distribution of OH. As discussed by \citet{sch2009}, the source of the low production observed in March 2008 is more likely the nucleus.

Assuming a nucleus bulk density equal to the dust density (700 kg m$^{-3}$, Sect~\ref{sec:model}), the mass ejected by the outburst is 3--9\% the nucleus mass, and corresponds to a cube having a side length of 0.8--1.2 km. In other words, the dramatic outburst experienced by 17P/Holmes was caused by a massive disruption of part of its nucleus.  Possible mechanisms for comet splitting are discussed by \citet{boen2005}. The simple fact that 17P/Holmes also experienced massive outbursts in 1892 and 1893 indicates that the outburst was not induced by a collision with another body. The idea that the outburst was triggered by the exothermic phase transition of buried amorphous water ice is defended by \citet{sekanina2008}, \citet{reach10}, and \citet{Li2011}. However, from thermal modelling of nucleus interior, \citet{kossacki2010} reach the conclusion that this process is unable to explain such a massive fragmentation. The origin of the fragmentation of comet 17P/Holmes remains elusive.      

\section{Summary}

The 3.3-mm continuum emission from comet 17P/Holmes was observed with the Plateau de Bure interferometer 4--5 days after its dramatic 2007 outburst. The main results of a detailed analysis of these observations are: 

\begin{itemize}
\item The peak position of the brightness distribution coincides within 1.5$\sigma$ with the nucleus position. Some excess emission is detected southward the nucleus position on  27 October.  This excess emission disappeared on  28 October. 

\item Two distinct dust components with different kinematics properties are identified in the data. 

\item The large-velocity component, with typical velocities $V_0$ of 50--100 m s$^{-1}$ for 1 mm particles, displays a steep size distribution with a size index estimated  to $q$ = --3.7 for the  nominal silicate-ice mixture assuming a minimum grain size of 0.1 $\mu$m  ($q$ within [--3.8, --3.6] considering a wide range of mixtures and porosities). It corresponds to the fast expanding shell observed in optical images. Velocities are consistent with acceleration by gas pressure resulting from the sublimation of icy grains. 

\item The slowly-moving component ($V_0$ = 7--9 m s$^{-1}$), referred to as the core, has a shallower size index $|q|$ $<$ 3.4, compared to the shell. These particles were not efficiently accelerated by gas drag. The dust mass in the core is in the range 0.1--1 that of the shell.   

\item  Using optical constants pertaining to porous grains (porosity of 0.5) made of astronomical silicates mixed with water ice (48\% in mass), the total dust mass $M_{\rm dust}$ injected by the outburst is estimated to 4--14 $\times$ 10$^{11}$ kg, corresponding to 3--9\% the nucleus mass. The dramatic outburst experienced by 17P/Holmes was caused by a massive disruption of its nucleus. 

\end{itemize}

\begin{acknowledgements}

We acknowledge the IRAM director P. Cox for scheduling these observations on a very short notice as a target of opportunity project and  J.-M. Winters for his helpful work during the observations. We thank J. Nicolas (IAU station B51) for providing us optical images of comet 17P/Holmes. The research leading to these results  received funding from the 
European Community's Seventh Framework Programme (FP7/2007--2013) under 
grant agreement No. 229517.
\end{acknowledgements}

\bibliographystyle{aa}
\bibliography{mnemonic,bigbiblio}

\end{document}